\documentclass[11pt]{article}

\usepackage{amsmath,amsthm,latexsym,amssymb,amsfonts,epsfig}

\numberwithin{equation}{section}
\allowdisplaybreaks

\usepackage{graphicx} 
\usepackage{physics}
\usepackage{color}

\usepackage[hidelinks]{hyperref}
\usepackage{booktabs}
\usepackage[compress]{cite}

\usepackage{bm}

\newtheorem{Definition}{Definition}
\newcommand{\RR}{\mathbb{R}}
\newcommand{\Fcal}{\mathcal{F}}
\newcommand{\Scal}{\mathcal{S}}

\title{{\sf Quantum Field Theory of}\\
{\sf Black Hole Perturbations with Backreaction}\\
{\sf VI. Apparent Horizons, Quasi-Local Mass and Effective Classical Metrics}}

\author{
{\sf J. Neuser}$^{1}$\thanks{{\sf neuser@qtc.sdu.dk}}, 
{\sf T. Thiemann}$^{2}$\thanks{{\sf
thomas.thiemann@fau.de}}\\
\\
{\sf $^1$ Quantum Theory Center ($\hbar$QTC) at IMADA \& D-IAS, Southern Denmark University}\\
{Campusvej 55, 5230 Odense M, Denmark}\\
{\sf $^2$ Institute for Quantum Gravity, FAU Erlangen -- N\"urnberg,}\\
{\sf Staudtstr. 7, 91058 Erlangen, Germany}\\
}
\date{{\small\sf \today}}

\begin{document}

\maketitle

\begin{abstract}
	In a recent series of papers we developed a first-principle and gauge invariant approach to black hole perturbation theory valid to any order.
	We included back reaction effects to tackle the situation of evaporating black holes and obtained an explicit expression for the dynamics of
	the reduced phase space to second order.
	The physics of evaporating black holes is in particular encoded by apparent horizons, an observer dependent generalisation of the event horizon.
	We determine the shape of the apparent horizon to second order in the perturbations.
	The area of the apparent horizon is an interesting observable which is expected to decrease in the quantum theory due to Hawking evaporation.
    We show how the full four dimensional metric can be reconstructed in terms of the reduced phase space variables.
    In the quantum theory, taking expectation values of this metric, we obtain an effective classical metric, whose causal structure can then be visualised 
    in a quantum corrected Penrose diagram.
    We conclude with an outlook into the quantisation procedure in the reduced phase space formalism and the implications on the area of the apparent horizon.
\end{abstract}

\section{Introduction}

In his seminal work, Hawking showed that in contrast to the picture of classical general relativity, black holes are not stable but emit radiation and thus 
lose energy and mass.
Semi-classical calculations suggest a black hole lifetime which scales like the cube of the initial black hole mass.
The change of the dynamics of the black hole spacetime due to the presence of the radiation is called backreaction.
So far, it has not been comprehensively derived using first principle calculations. 
In particular the final phase of the evaporation process is still under large debates as the semi-classical approximation is expected to break down. 

In \cite{I,II,III,IV,V}, we developed a novel, concrete and gauge-invariant formulation of quantum field theories on spherically symmetric spacetimes including
geometry -- geometry and matter -- geometry back reaction. 
We explored in particular scalar and electromagnetic matter but the formalism is capable to treat all matter types
of the standard model.  
Motivated by the spherical symmetry of the Schwarzschild solution, we split the degrees of freedom into spherically symmetric and non-symmetric degrees of freedom.
The symmetric degrees of freedom are the background and the non-symmetric degrees of freedom the perturbations. 
On the full phase space we further split the degrees of freedom into observable and gauge degrees of freedom. 
This construction is manifestly non-perturbative and valid to all orders in the non-symmetric variables. 
We obtain a description in terms of four disjoint sets of variables (symmetric observable, symmetric unobservable, non-symmetric observable and non-symmetric unobservable).
All the physics is contained in the dynamics of the observable degrees of freedom.

For the dynamics of the theory, we introduce the reduced phase space, coordinatised by the observable degrees of freedom.
Then, the time evolution on the reduced phase space is generated by the physical Hamiltonian, known implicitly to any order in perturbation theory. 
We calculated it explicitly to second and third order in the non-symmetric degrees of freedom \cite{II,III,IV,V}.
The results agree with previous calculations in the literature when we drop back reaction terms 
and transform to the correct coordinate system \cite{1,2,6}. The latter step is necessary because we work in Gullstrand-Painleve 
coordinates rather than the Schwarzschild coordinates mostly used in the literature in order to be able to cover also 
the black hole interior.

In this manuscript we apply our approach to physical predictions 
relevant for black hole evaporation by discussing suitable evaporation sensitive observable quantities in the reduced phase space
language. We focus on the vacuum and electro vacuum case but the extension to other matter is straightforward. 
In \cite{I}, a concrete such observable, the apparent horizon \cite{HawkingEllis,AnninosApparent,AshtekarDynamical,DeformedHorizon},
was analysed within our framework. 
The apparent horizon is a quasi-local and foliation dependent generalisation of the event horizon.
The foliation is motivated by considering an observatory studying an evaporating black hole.
The closest an observatory can be to an inertial frame of reference is if it is freely falling in the black hole spacetime. 
Physically, the time a radially falling observer in a black hole spacetime would measure is the Gullstrand-Painleve (GP) time. 
The corresponding GP coordinates, which are in this sense distinguished on physical grounds,  
are naturally associated to such an observer and we choose the corresponding GP foliation. 
As an additional benefit, the GP coordinates are regular across the horizon and cover both the interior and exterior
of the black hole.

As we will demonstrate in the following sections and as we proved abstractly in \cite{I}, 
it is possible to explicitly determine the shape of the apparent horizon perturbatively.
This gives us direct access to the change of the shape of the black hole due to  back reaction, i.e. the  
influence of non-symmetric degrees of freedom on symmetric observables. 
The area of the apparent horizon is of particular interest because it is the property of a local
quantity in contrast to the event horizon which is a global quantity and thus accessible to 
local observations. 
In fact we consider also the square root of the area
which introduces the notion of a quasi-local mass, which can serve as the dynamical mass during the Hawking process.

Classical General Relativity proves within the usual assumptions 
(e.g. energy conditions, global hyperbolicity) the black hole area theorem, namely 
that the area of the event horizon can never shrink and 
that the apparent horizon is always inside the event horizon. Therefore, 
if a black hole evaporates as the existence of Hawking radiation suggests, then 
in a semiclassical sense at least one of the assumptions of the area theorem must 
be violated. A possible mechanism for such violations are fluctuations in quantum field 
theories which e.g. can lead to violations of classical energy inequalities \cite{Fewster}. 
This suggests the following semiclassical process: Working in the Heisenberg picture, 
the local measurement of a 
quantum spacetime metric in the vicinity of a hypersurface by taking expectation values with 
respect to a semiclassical (say minimal uncertainty) state delivers initial data. 
The evolution of those initial data from that hypersurface using the classical Einstein equations  
will generically generate a globally hyperbolic spacetime that generically contains an event horizon. 
However, the true evolution of the spacetime is not using the classical Einstein equations but 
the quantum Einstein equations. Therefore, if the above mechanism is at work, measuring 
the spacetime metric on a later hypersurface with respect to the classical spacetime predicted by 
the first measurement will yield new initial data that predict an event horizon with smaller 
area. We are pedantic about this point, because if one just relies on a strictly 
classical picture and if the black hole is eventually absent due to evaporation then 
since the event horizon is a global concept, it should not have been there in the first place and 
thus there could never be an apparent horizon. However, within the semiclassical process just sketched 
the existence of a non-empty apparent horizon before evaporation is not a contradiction and 
its area must eventually shrink to zero.    

In order to monitor this mechanism in full detail, it is important to have access to the 
full spacetime metric as a quantum object expressed in terms of gauge invariant quantum 
degrees of freedom. The reduced phase space formalism is capable of precisely accomplishing that
(see also \cite{ReferenceFrame} and references therein for more details):
1. A choice of reference fields (here chosen as a subset of the spatial metric components to 
fix the Gullstrand-Painleve gauge) and 
gauge conditions on them defines a reference frame. 2. The constraints are solved for the 
momenta canonically conjugate to the reference fields. 3. The stability of the gauge condition
under time evolution fixes lapse and shift in terms of the remaining, so-called true, degrees 
of freedom. 4. The evolution of the true degrees of freedom in the chosen reference frame, 
on the constraint surface and with respect to the fixed lapse and shift is described by
a reduced Hamiltonian. 5. The true degrees of freedom and its reduced Hamiltonian 
is in 1-1 correspondence with corresponding gauge invariant Dirac observables and a physical 
Hamiltonian so that one can identify the reduced and manifestly gauge invariant descriptions.  

In the previous works in this series of papers, we focussed on the perturbative construction of 
the physical Hamiltonian and the solutions of the constraints to second and third order but 
lapse and shift were not considered so far. Thus in this paper we fill this gap
and derive the solution for lapse and shift in terms of the observable degrees of freedom to second order
(the third and higher order can also be calculated systematically but will be reserved for future work, see 
appendix).
This gives access to the full four-dimensional metric in terms of the observable, true degrees of freedom
as outlined above. The resulting expressions can now be quantised in suitable (Fock) representations 
that are motivated by the the structure of the physical Hamiltonian. Then we can follow 
the above process and monitor the expectation value of the metric components in the 
Heisenberg picture generated by the reduced quantum Hamiltonian. This gives rise 
to a dynamical, semiclassical spacetime whose Penrose diagram can be computed and 
in which the apparent horizon can be located.\\ 
\\
We organise this work as follows:\\

In section \ref{sec:RevPert} we review the derivation of the second order reduced Hamiltonian providing all necessary formulas for later chapters.

In section \ref{sec:ApparentHorizon} we define the apparent horizon, perform the explicit computations and discuss the physical implications.

In section \ref{sec:LapseShift}, we derive expression for lapse and shift in terms of the reduced phase space variables. 

In section \ref{sec:Penrose} we explore possible approaches to the computation of semiclassical Penrose diagrams
as outlined above. This step is not yet complete. 

In section \ref{sec:Conclusion} we summarise and conclude.
 
In appendix \ref{sec:EvenParityBdryEM}, we collect a boundary term for the even polarity perturbations and in appendix 
\ref{sec:ThirdOrder} we show that the analysis for the apparent horizon can be extended to third order in the perturbations.

\section{Second Order Black Hole Perturbation Theory}
\label{sec:RevPert}

In this section, we briefly review the reduced phase space formulation of black hole perturbation theory.
The result is a physical Hamiltonian which generates the time evolution of the observables coordinatising the reduced phase space. 
Furthermore, we summarise the solution of the constraints which will be used in later chapters.
More details on the derivation can be found in \cite{I,II}.

The core of the reduced phase space approach is a splitting of the degrees of freedom into four disjoint sets:
First, we use spherical symmetry to distinguish between the symmetric and non-symmetric degrees of freedom.
The symmetric degrees of freedom describe the background (e.g. the mass of the black hole) while the non-symmetric degrees of freedom represent gravitational waves which dynamically interact with the background. 
Since general relativity is a fully constrained system, we need to address the issue of gauge invariance to extract physical predictions.
In our formulation, we achieve this by distinguishing between true (observable) and gauge (unobservable) degrees of freedom.
The momenta of the gauge degrees of freedom are fixed by solving the constraint equations and the configuration variables are pure gauge and can be fixed using gauge fixing conditions.
Then, the physical content is described by the reduced phase space, coordinatised by the observable degrees of freedom.
Finally, the dynamics is described by the reduced/physical Hamiltonian which we derived to second order in the non-symmetric degrees of freedom in 
\cite{I,II}. For more information about this procedure and why it is manifestly gauge invariant when properly interpreted despite fixing a gauge 
see \cite{ReferenceFrame}.

We work in ADM variables, i.e. the spatial metric $m_{ij}$ and its conjugate momentum $W^{ij}$.
In \cite{II}, we split the variables into radial and angular components and expand them in terms of spherical harmonics. 
For the gauge fixing condition we decided to use the Gullstrand-Painlevé gauge ($m_{33}=1$, $m_{3A}=0$ and $\Omega^{AB}m_{AB} = 2 r^2$)
because it is regular across the horizon. 
In this gauge, we adopt the following notation for the components of the metric and its conjugate momentum:
\begin{align}
\begin{aligned}
    m_{33} &= 1, \quad W^{33} = \sqrt{\Omega}p_v + \sqrt{\Omega} \sum_{l\geq 1, m}  y_v^{lm} L_{lm}\\
    m_{3A} &= 0, \quad W^{3A} = 0 +  \sqrt{\Omega} \frac{1}{2}\sum_{l \geq 1, m,I} y_I^{lm} L_{I,lm}^A\\
    m_{AB} &= r^2 \Omega_{AB} + \sum_{l \geq 2, m, I}  X^I_{lm} L^{I,lm}_{AB}\\
    W^{AB} &= \sqrt{\Omega} \frac{p_h}{2} \Omega^{AB} + \sqrt{\Omega} \frac{1}{2}\sum_{l\geq 1,m} 
     y_h^{lm} \Omega^{AB} L_{lm} + \sqrt{\Omega} \sum_{l \geq 2, m, I}  Y_I^{lm} L_{I,lm}^{AB}
\end{aligned}
\end{align}
The indices $l=0,1,2,\dots$ and $m=-l,\dots,l$ are the usual spherical harmonic indices and $I$ runs over the even 
or polar ($I=e$) and odd or axial ($I=o$) vector and tensor harmonics.
The symmetric degrees of freedom are $p_v, p_h$ and $y,X,Y$ are the non-symmetric degrees of freedom.
The gauge momenta are $p_v$, $p_h$ in the symmetric sector as well as $y_v, y_h$ and $y_e$ in the non-symmetric sector. 
In the present section we consider only the gravitational degrees of freedom. The extension to Maxwell matter is 
reviewed in the next section.

For a well defined Hamiltonian formulation, we need to provide fall-off conditions on the variables.
These conditions need to be chosen such that i. the Hamiltonian and the symplectic structure have well defined integrals and
ii. the variational derivatives are well defined. 
For the second point it is sometimes necessary to introduce additional (counter) boundary terms to the bulk Hamiltonian. 
In the present paper, we work in the reduced phase space formulation requiring a more careful choice of fall-off conditions \cite{ThiemannDecayConsistent}.
The decay conditions need to be consistent with the solution of the constraints for the gauge degrees of freedom.
In that way the decay conditions of the gauge momenta is constrained by the decay conditions of the physical degrees of freedom. 
Additionally, the requirement that the gauge fixing is preserved under time evolution, leads to stability conditions which fix lapse and shift and constrains their decay behavior.
Then, one needs to ensure that everything is consistent: i. the smeared constraints converge for the specified decay conditions,
ii. the variation of the constraint can be written as a bulk integral with a potential boundary term given by a total differential,
iii. the boundary term converges. 
In \cite{ThiemannDecayConsistent} a general strategy for obtaining such decay conditions is outlined.
It is explicitly applied to the present situation in \cite{V}. 
We find the decay behavior, $p_v = O(\sqrt{r})$ and $p_h = O(r^{-3/2})$ for the symmetric variables and 
$y_v = O(r^{1/4})$, $y_h = O(r^{-7/4})$, $y_e = O(r^{-3/4})$.
For the physical degrees of freedom, we found  $(X^o, Y_o) = (O(r), O(r^2)$ and $(X^e, Y_e) = (O(r^{3/4}), O(r^{7/4})$.

At zeroth order we found the solution for the background momenta as
\begin{equation}
    p_v^{(0)} = 2 \sqrt{r r_s} \quad p_h^{(0)} = r^{-2} \sqrt{r r_s}\,,
\end{equation}
Here, $r_s = 2M$ is the Schwarzschild radius, where $M$ is the Schwarzschild mass which arises as an integration constant.
The above solution for the background momenta corresponds to the Schwarzschild solution in Gullstrand-Painlevé coordinates.

To first order, the constraints split into the odd and even polarity contributions and we solve them separately.
For the odd polarity sector we only have the axial angular diffeomorphism constraint.
Its solution for $y_o^{(1)}$ is 
\begin{align}
    y_o^{(1)} = \frac{\sqrt{2(l+2)(l-1)}}{r^2}\int \dd{r} \qty(r^2 Y_o + \frac{p^{(0)}_h}{2} X^o)
\end{align}
The physical Hamiltonian simplifies drastically when expressed in so-called master variables introduced using a canonical transformation.
The odd polarity master variables $Q^o, P_o$ are defined as
\begin{align}
    Q^o &:= \frac{\sqrt{2}}{r} \int \dd{r} \qty(r^2 Y_o + \frac{p^{(0)}_h}{2}X^o)\\
    P_o &:= \frac{r}{\sqrt{2}}\partial_r \qty(r^{-2} X^o) + \frac{p_v^{(0)}}{2r^2} Q^o
\end{align}
It is convenient to write the canonical transformation in terms of a generating functional of type 1 $G(X^o, Q^o)$ such that
\begin{equation}
    Y_o = \fdv{G}{X^o}, \quad P_o = - \fdv{G}{Q^o}
\end{equation}
Explicitly, we find 
\begin{equation}
    G = \int \dd r\qty[\frac{1}{r} (r Q^o)' \tilde X^o - \frac{\sqrt{r r_s}}{2 r^2}((\tilde X^o)^2 + (Q^o)^2)]
\end{equation}
where we defined $X^o := \sqrt{2} r \tilde X^o$ and $Y_o := 1/(\sqrt{2} r) \tilde Y_o$.

We continue with the solution of the first-order even polarity constraints.
There, we solve the polar angular diffeomorphism constraint, the radial diffeomorphism constraint and the Hamiltonian constraint for the momenta $y_h, y_v$ 
and $y_e$. 
The solutions for $y_h$ and $y_e$ are given in terms of $y_v$ as
\begin{align}
    \label{eq:yeyh}
	\sqrt{l(l+1)} y_e^{(1)} &=2 \partial_r y_v^{(1)} - \frac{1}{r} y_v^{(1)} + \frac{\sqrt{2(l+2)(l+1)l(l-1)}}{2 r \sqrt{r r_s}} X^e\\
    y_h^{(1)} &= \frac{y_v^{(1)}}{2r^2} - \frac{\sqrt{2(l+2)(l+1)l(l-1)}}{4 r^2 \sqrt{r r_s}} X^e
\end{align}
$y_v^{(1)}$ is the solution of a differential equation and the particular solution is implicitly given by an integral
\begin{align}
    y_v^{(1)} = r^{\alpha_-}\int \dd{\tilde r}\qty(\frac{2 i \tilde r^{\alpha_+-\frac{1}{2}}}{\sqrt{4l(l+1)-9}} s(\tilde r)) - r^{\alpha_+}\int \dd{\tilde r}\qty(\frac{2 i \tilde r^{\alpha_- -\frac{1}{2}}}{\sqrt{4l(l+1)-9}} s(\tilde r))
    \label{eq:yv}
\end{align}
where
\begin{equation}
    \alpha_\pm = \frac{1}{4}\qty(-1 \pm i \sqrt{4l(l+1) - 9})
\end{equation}
and
\begin{equation}
    s(r):=- \sqrt{2(l+2)(l+1)l(l-1)}\qty(r^2  Y^e + \frac{p_h^{(0)}}{2}  X^e + \frac{l(l+1) - 1}{2 \sqrt{r r_s}}  X^e - \frac{r}{\sqrt{r r_s}} \partial_r  X^e)\,.
\end{equation}

Analogously, we introduce master variables for the even polarity sector to simplify the physical Hamiltonian. We define $Q^e,P_e$ using a canonical 
transformation given by
\begin{align}
    Q^e &:= \frac{r + r_s}{\Lambda}\sqrt{\frac{(l+2)(l-1)}{l(l+1)}} q_1\\
    \begin{split}
    P_e &:= \frac{\Lambda}{r+r_s}\sqrt{\frac{l(l+1)}{(l+2)(l-1)}} p_1 - \frac{\sqrt{r r_s}}{{8 r^3 (r+r_s) \Lambda^2}} \Big(6 \left(5 l^2+5 l-31\right) r_s^3\\
    &+(l+2)^2(l-1)^2 \left(l^4+2 l^3+13 l^2+12
   l+3\right) r^3+3 \left(9 l^4+18 l^3+5 l^2-4 l+35\right) r r_s^2\\
   &+2 (l+2)(l-1)(l^2 + l + 1) \left(5 l^2 + 5 l + 18\right) r^2 r_s \Big) \frac{q_1}{\sqrt{(l+2)(l+1)l(l-1)}}
   \end{split}
\end{align}
where 
\begin{equation}
    \Lambda := \frac{1}{2}(l+2)(l-1) + \frac{3}{2} \frac{r_s}{r}
\end{equation}
and introduced new quantities $q_1$ and $p_1$ related to the original variables via
\begin{align}
	q_1 :=& \frac{\sqrt{r r_s} }{r(r+r_s)} y_v^{(1)} + \sqrt{\frac{l(l+1)}{2 (l+2)(l-1)}}\frac{\Lambda}{r^2 (r+r_s)} X^e\\
    \begin{split}
	    p_1 :=&  y_v^{(1)} - \sqrt{r r_s}\frac{\qty(3 -l(l+1) (l^2+l+10) )r - 2 (l^2+l-9) r_s}{2 \sqrt{2 (l-1) l (l+1) (l+2)} r^2 (r+r_s)}  X^e\\
    &-\partial_r\qty(\frac{2 r^2 (r+r_s)}{l (l+1) \Lambda}p_2) + \frac{r^2 \qty((l^2 + l + 2) - 3 \frac{r_s}{r})}{l (l+1) \Lambda}p_2
    \end{split}
\end{align}
The variable $p_2$ is defined as
\begin{align}
	p_2 :=&  y_h^{(1)} + \frac{1}{2 r (r + r_s)}\qty(\frac{r_s}{r} - (l(l+1)+2))  y_v^{(1)} - \partial_r (r^{-1}  y_v^{(1)})
\end{align}
As in the odd polarity case, we introduce a type 1 generating functional $G(\tilde X^e, Q^e)$ satisfying
\begin{align}
    \begin{split}
    \fdv{G}{X^e} = Y_e &= - \pdv{r}\qty(\frac{2 r \Lambda}{(l+2)(l-1)} \sqrt{\frac{r}{r_s}} (\tilde X^e)') - \frac{1}{4 r \sqrt{r r_s}}\qty(3(l+2)(l-1)r + 7 r_s)\tilde X^e\\
    &+\frac{1}{r \Lambda} \pdv{r}\qty(\frac{2 r^2 \Lambda^2}{(l+2)(l-1)}\sqrt{\frac{r}{r_s}}(Q^e)') + \frac{1}{4 r \sqrt{r r_s}}\qty(l(l+1)r + 3 r_s)Q^e
    \end{split}\\
    \begin{split}
    - \fdv{G}{Q^e} = P_e &= \partial_r\qty(\frac{2 r \Lambda}{(l+2)(l-1)}\sqrt{\frac{r}{r_s}} \partial_r Q^e) - \frac{(l+2)(l+1)l(l-1)r^2 - 6 r r_s + 3 r_s^2}{8 r^2 \sqrt{r r_s} \Lambda} Q^e\\
    &- \frac{2 r \Lambda}{(l+2)(l-1)}\sqrt{\frac{r}{r_s}} \partial_r^2 \tilde X^e - \frac{(l+2)(l-1)r + 3 r_s}{2(l+2)(l-1)\sqrt{r r_s}} \partial_r \tilde X^e - \frac{(l+2)(l-1)r + 3 r_s}{4 r \sqrt{r r_s}} \tilde X^e
    \end{split}
\end{align}
where $X^e := \sqrt{2} r \tilde X^e$ and $Y_e := 1/(\sqrt{2} r) \tilde Y_e$.
Using Mathematica and the canonical transformation above we find the result
\begin{align}
    G &= \int \dd r \Big[\frac{r \Lambda}{(l+2)(l-1)} \sqrt{\frac{r}{r_s}} \qty((\partial_r \tilde X^e)^2 + (\partial_r Q^e)^2) - \frac{1}{8 r \sqrt{r r_s}}\qty(3(l+2)(l-1)r + 7 r_s)(\tilde X^e)^2\\
    &+\frac{(l+2)(l+1)l(l-1)r^2 - 6 r r_s + 3 r_s^2}{16 r^2 \sqrt{r r_s} \Lambda} (Q^e)^2\\
    &- \frac{2 r^2 \Lambda^2}{(l+2)(l-1)}\sqrt{\frac{r}{r_s}}(Q^e)' \pdv{r}\qty(\frac{\tilde X^e}{r \Lambda}) + \frac{1}{4 r \sqrt{r r_s}}\qty(l(l+1)r + 3 r_s)Q^e \tilde X^e \Big] 
\end{align}

Next, we determine the solution of the second order symmetric constraints: the radial diffeomorphism and the Hamiltonian constraints.
The solutions give second order contributions $p_v^{(2)}$ and $p_h^{(2)}$ to the symmetric gauge momenta, which depend quadratically on $X,Y$. 
We find
\begin{align}
    p_v^{(2)} &= \frac{1}{8\pi} \sqrt{\frac{r}{r_s}}\qty(\int E \dd{r} + B_o + B_e)\\
    p_h^{(2)} &= \frac{p_v}{2 r^2} + \frac{1}{8 \pi \sqrt{r r_s}}{}^{(2)}C_v\,.
\end{align}
The solution is written in terms of an integral over the radial density $E(r)$ and the boundary terms $B_o$ ($B_e$) for the 
even (odd) polarity respectively.
$E(r)$ is given by
\begin{equation}
    E(r) = \sum_{I,lm} \sqrt{\frac{r_s}{r}}P_{I,lm} \partial_r Q^{I,lm} + \frac{1}{2}\qty((P_{I,lm})^2 + (\partial_r Q^{I,lm})^2 + V_I (Q^{I,lm})^2)\,.
\end{equation}
Again, the index $I$ runs over the even and odd polarity contributions and $V_I$ are the Regge-Wheeler-Zerilli potentials defined as
\begin{align}
    V_o &= \frac{1}{r^3}\qty(l(l+1)r - 3 r_s)\\
    V_e &= \frac{l(l+1)(l+2)^2(l-1)^2 + 3 (l+2)^2(l-1)^2 \frac{r_s}{r} + 9 (l+2)(l-1) \frac{r_s^2}{r^2} + 9 \frac{r_s^3}{r^3} }{4 r^2 \Lambda^2}
\end{align}

The boundary terms have also been worked out in the previous paper \cite{II}, and we showed that they vanish at infinity.
For the computations in this paper we need their values for finite $r$, so they can no longer be neglected.
In their simplified form and in GP gauge they are given by
\begin{align}
	B_o = \sum_{lm}\qty[ \frac{r-r_s}{r^4} \qty[\int \dd{r}\qty(r^2  Y_{o,lm} + \frac{p_v}{2}  X^o_{lm})]^2 + \frac{6 r - r_s}{4 r^4} (X^o_{lm})^2 - \frac{1}{r^2}  X^o_{lm} (X^o_{lm})' ]
\end{align}
and
\small
\begin{align}
	B_e &= \sum_{lm} \left[\frac{(l+2)(l-1)r^2+ 5 (l+2)(l-1) r r_s - (l^2+l-17)r_s^2}{8 r^4 r_s \Lambda} X^e_{lm} X^e_{lm} - \frac{1}{r^2} X^e_{lm} ( X^e_{lm})'\right.\nonumber \\
	    &+ \sqrt{\frac{2(l+2)(l-1)}{l(l+1)}}\frac{r-r_s}{2 r \sqrt{r r_s} \Lambda}  X^e_{lm} \partial_r  y_{v,lm}^{(1)} + \frac{2 r (r-r_s)}{2 l(l+1) r \Lambda}\partial_r  y_{v,lm}^{(1)} \partial_r y_{v,lm}^{(1)} + \frac{2((l^2+l-1)r + r_s)}{2 l(l+1) r \Lambda}  y_{v,lm}^{(1)} \partial_r  y_{v,lm}^{(1)} \\
    &+\sqrt{\frac{(l+2)(l-1)}{2l(l+1)}}\frac{(l-1) (l+2) \left(l^2+l-1\right) r^3+\left(l(l+1) \left(l^2+l+2\right)-5\right) r^2 r_s+3 \left(l^2+l-1\right) r
    r_s^2+6 r_s^3}{r^{5/2} \sqrt{r_s} \left(\left(l^2+l-2\right) r+3 r_s\right)^2}  X^e_{lm} y_{v,lm}^{(1)}\nonumber\\
   &\left.+\frac{\left(l^2+l-2\right)^2 r^3+(l-1) (l+2) \left(l(l+1) \left(l^2+l-3\right)+8\right) r^2 r_s +3 \left(l^2+l-11\right)
   r_s^3-9 (2 l (l+1)-5) r r_s^2}{16 l (l+1) r^4 \Lambda^3}  (y_{v,lm}^{(1)})^2\right] \nonumber
\end{align}
\normalsize

The Hamiltonian analysis in \cite{I}, which is based on a careful study of fall-off conditions and boundary terms, leads to an implicit expression for the 
physical Hamiltonian to all orders in the perturbations.
For one asymptotic end and to second order, in \cite{II}, we obtained the result
\begin{align}
    H_\mathrm{phys} = M + \frac{1}{\kappa} \int_\mathbb{R^+} E \dd{r}\,.
\end{align}
where $\kappa$ is the gravitational coupling constant.

\section{Apparent Horizon and Quasi-local Mass}
\label{sec:ApparentHorizon}

The notion of event horizon in general relativity is a global concept which requires the knowledge of the global spacetime. 
The event horizon is defined as the boundary of the region which is not in the past of future null infinity.
In contrast to that, the apparent horizon is a more local concept. 
It is defined using the notion of trapped regions.
In the beginning of this chapter, we review important definitions following the notation in \cite{I}.

After the introduction of the general formalism, we apply it to the case of black holes in Gullstrand-Painleve coordinates. 
In the first part, we discuss the purely gravitational case and 
generalise this in the next section to the Einstein-Maxwell system.
In both cases, we determine the shape of the apparent horizon and the quasi-local mass.
We conclude with a brief discussion on the time-dependence of the quasi-local mass and its relation to the flux through the horizon.

\subsection{Apparent Horizons in General Relativity}

The spacetime $(M,g)$ is assumed to be globally hyperbolic and foliated into three-dimensional spatial submanifolds.
Let $\Sigma$ be a Cauchy surface with timelike normal $n$ ($g(n,n)=-1$).
Furthermore, we consider a closed, oriented 2-surface $S$ in $\Sigma$ without boundary.
This surface has a spacelike normal $s$ which is tangential to $\Sigma$, i.e. $g(n,s)=0$ and $g(s,s)=1$.
The induced metric on the Cauchy surface is given by $m=g + n \otimes n$ and the induced metric on $S$ is given by $h=m - s\otimes s$. 

We then define the future outward/inward oriented null vectors $l_\pm = n \pm s$.
At each point $y^A$ ($A=1,2$) on $S$ we construct affinely parametrized geodesic $c_{y,\pm}(\lambda)$ with initial tangent $l_\pm$
and parameter $\lambda$.
The $c_{y,\pm}$ define two null geodesic congruences $C^{\pm}_S$ with tangent vectors 
$\partial_\lambda^\pm = \pdv{c^\mu_{y,\pm}(\lambda)}{\lambda} \partial_\mu =: l_\pm$ 
and deviation vectors $\partial_A^\pm = \pdv{c^\mu_{y,\pm}(\lambda)}{y^A} =: e_{A,\pm}$.

In \cite{I}, it is reviewed that the full content of the geodesic deviation is contained in the quantity 
\begin{equation}
    \kappa_{AB}^{\pm} = g(e_{A,\pm},\nabla_{e_{B,\pm}} l_{\pm})
\end{equation}
Furthermore, on $C^{\pm}_S$ we introduce 
\begin{equation}
    h_{AB}^\pm = g(e_{A,\pm},e_{B,\pm})\,.
\end{equation}
and its inverse $h^{AB}_\pm$. Then, we decompose $\kappa_{AB}^\pm$ into expansion, shear and rotation:
\begin{align}
    \theta_{\pm} := h^{AB}_\pm \kappa_{AB}^\pm, \quad \sigma_{AB}^\pm := \kappa^\pm_{(AB)} - \frac{1}{2} \theta_{\pm} h^\pm_{AB}, \quad \omega_{AB}^\pm := \kappa^\pm_{[AB]}
\end{align}

The expansion is of special interest in the following discussion. 
It captures the divergence and convergence of the geodesic congruences. 
In the notation introduced above, we derive an explicit formula for it:
\begin{align}
\begin{split}
    \theta_\pm &= h^{i j}_\pm \nabla_i l_{\pm j} = (m^{i j} - s^i s^j) \nabla_i(n_j \pm s_j)\\
    &= K - s^i s^j K_{ij} \pm m^{ij} \nabla_i s_j\\
    &=- s^i s^j \frac{W_{i j}}{\sqrt{m}} \pm D_i s^i
\end{split}
\label{eq:thetaPlus}
\end{align}
where $i,j=1,2,3$ are spatial indices for the coordinates $y^3=r, y^A,\; A=1,2$ and $D$ is the covariant 
differential with respect to $m$. 
Multiplying with the square root of the determinant of $m$ we can rewrite the covariant derivative as a partial one and obtain
\begin{align}
    \sqrt{m}\theta_\pm = - s_i s_j W^{ij} \pm \partial_ i (\sqrt{m} m^{ij} s_j)\label{eq:ApparentHorizon}
\end{align}

We recall the following definitions for trapped surfaces, apparent horizons and quasi-local masses:

\begin{Definition}
Consider a globally hyperbolic spacetime $(M, g)$ and a Cauchy surface $\Sigma$ in it.
\begin{enumerate}
    \item A closed, orientable 2-surface $S \subset \Sigma$ without boundary $\partial_\Sigma S= \emptyset$ is called trapped if $\theta_+ = 0$.
    \item A trapped region in $\Sigma$ is a closed subset $T \subset \Sigma$ such that $S := \partial_\Sigma T$ is trapped.
    \item The trapped surface in $\Sigma$ defined by the total trapped region (closure of union of all trapped regions) is called the apparent horizon $A_\Sigma$ of $\Sigma$.
\end{enumerate}
\end{Definition}
\begin{Definition}
Consider a globally hyperbolic spacetime $(M, g)$ and a foliation $\Fcal = \bigcup_{\tau \in \RR}\Sigma_\tau$ of $M$ by Cauchy surfaces $\Sigma_\tau$.
\begin{enumerate}
    \item If $\tau \mapsto S \subset \Sigma_\tau$ is a one parameter family of trapped surfaces then $\Scal := \bigcup_{\tau \in \RR} S_\tau$ is called a trapping horizon.
    \item Let $A_\tau := A_{\Sigma_\tau}$ be the apparent horizon of $\Sigma_\tau$. Then $A_\Fcal := \bigcup_{\tau \in \RR} A_\tau$ is called the apparent horizon of $\Fcal$.
\end{enumerate}
\end{Definition}

\begin{Definition}
Given a foliation $\Fcal$ of a globally hyperbolic $(M, g)$ by Cauchy surfaces $\Sigma_\tau$ the quasi-local mass at time $\tau$ is defined as
\begin{equation}
    [M_0]^2 := \mathrm{Ar}[A_\tau]
\end{equation}
i.e. the square root of the apparent horizon area.
\end{Definition}

\subsection{Purly Gravitational Case}

In \cite{I}, it was shown that, using parts of the 
technology developed in \cite{DeformedHorizon} within in the Lagrangian formulation,
the apparent horizon can be found perturbatively order by order 
entirely within the canonical formalism. We now perform the corresponding computations explicitly.
As we will see, the apparent horizon on a fixed time slice 
of the purely spherically symmetric problem is a sphere at the Schwarzschild radius $r_s$.
This motivates us to consider embeddings of 2-surfaces of spherical topology to define the apparent horizon 
in the presence of non-symmetric perturbations.
We assume the deviation from the Schwarzschild horizon to be small, so that perturbation theory applies. 

The apparent horizon is a hypersurface and we choose the embedding $Y: \RR \times S^2 \to M$
with $Y(\tau,\cdot): S^2 \to \Sigma_\tau$ to describe it.
We assume the embedding to be of the form
\begin{align}
    Y^\tau(\tau, y) = \tau,\quad Y^3(\tau,y) = \rho(\tau,y), \quad Y^A(\tau, y) = y^A\,.
 \end{align}
 It is parametrized by a function $\rho(\tau,y)$. This function is called the \emph{radial profile} and describes the shape of the surface.

The embedding defines tangent vectors $Y_{(A)}$. Using differentiation along the angular coordinates we find 
\begin{equation}
    Y_{(A)} = \partial_A + D_A \rho \partial_r
\end{equation}
Since the embedding is of co-dimension 1 in $\Sigma$, we obtain the non-normalised co-normal $\hat s$ to the surfaces defined by $Y$ as
\begin{align}
    \hat s = \frac{1}{2}\epsilon_{ijk} Y^i_{(A)} Y^j_{(B)} \epsilon^{AB} \dd{x}^k =\dd{r} - (D_A \rho) \dd{x}^A
\end{align}
The norm of it is computed using the inverse metric $m^{-1}$ and in GP gauge ($m_{33}=1$, $m_{3A}=0$, $\Omega^{AB} m_{AB} = 2 r^{2}$) we have
\begin{equation}
    m^{-1}(\hat s, \hat s) = m^{33} \hat s_3 \hat s_3 + m^{AB} \hat s_A \hat s_B = 1 + m^{AB} D_A \rho D_B \rho\,.
\end{equation}
Therefore, the normalized co-normal $s$ is given by
\begin{equation}
    s = \frac{\hat s}{\sqrt{1 + m^{AB} D_A \rho D_B \rho}} \sim \qty(1 - \frac{1}{2} m^{AB} D_A \rho D_B \rho)\hat{s}\,,
\end{equation}
where we expanded the square root to second order in the perturbations. We assumed that the non spherically symmetric part of $\rho$ is at least of first order.

In Section \ref{sec:RevPert}, we introduced the decomposition of the metric into spherical symmetric and non-symmetric variables. 
We have $m_{33}=1$, $m_{3A} = 0$ and $m_{AB} = r^2 \Omega_{AB} + X_{AB}$, where for now we did not expand in terms of spherical harmonics.
The inverse metric to second order is $m^{33}=1$, $m^{3 A} = 0$ and $m^{AB} \sim r^{-2}\Omega^{AB} - r^{-4}  X^{AB} + r^{-6}  X^{AC}  X^{BD}\Omega_{CD}$.
The conjugate momentum is decomposed as $W^{33} = \sqrt{\Omega}( p_v +  y_v)$, $W^{3A} = \frac{\sqrt{\Omega}}{2}  y^A$, $W^{AB}= \sqrt{\Omega}(p_h +  y_h)\Omega^{AB}/2 + \sqrt{\Omega}  Y^{AB}$. 
For the computation we also require the square root of the determinant of the induced metric $m$. We have
\begin{align}
    \sqrt{\det m} = \sqrt{\det(r^2 \Omega+X)} = \sqrt{r^4 \det \Omega + \det X} \sim r^2 \sqrt{\det\Omega}\qty(1 - \frac{1}{4r^4} X^{AB} X_{AB})\,.
\end{align}

We now have all the ingredients to evaluate the condition of a vanishing outward null expansion $\theta_+$. 
The condition is given by setting \eqref{eq:thetaPlus} equal to zero. The first term involving the momentum $W^{ij}$ gives
\begin{align}
    \begin{split}
    - s_i s_j W^{ij} &= - s_3 s_3 W^{33} - 2 s_3 s_A W^{3A} - s_A s_B W^{AB}\\
    &\sim \sqrt{\Omega}\qty[- (1 - m^{AB} D_A \rho D_B \rho) (p_v + y_v) + D_A \rho y^A - \frac{1}{2}D_A\rho D_B \rho \Omega^{AB} p_h]
    \end{split}
\end{align}
The second term in the condition is given by the divergence of a vector density.
We expand the vector density to second order and obtain
\begin{align}
    \sqrt{m} m^{33} s_3 &\sim r^2 \sqrt{\Omega}\qty(1 - \frac{1}{4r^4} X^{AB} X_{AB} - \frac{1}{2 r^2} \Omega^{AB} D_A \rho D_B \rho)\\
    \sqrt{m} m^{AB} s_B &\sim \sqrt{\Omega}(- D^A \rho + r^{-2} X^{AB} D_B \rho)
\end{align}
Taking the radial and angular derivatives respectively, we have
\begin{align}
    \partial_r(\sqrt{m} m^{33} s_3) &\sim \sqrt{\Omega}\qty(2 r - \frac{1}{2} X^{AB}/r \partial_r (X_{AB}/r))\\
    \partial_A(\sqrt{m} m^{AB} s_B) &\sim - \sqrt{\Omega} D_A D^A \rho + r^{-2} \sqrt{\Omega} D_A(X^{AB} D_B \rho)
\end{align}
Then, combining the terms, the trapping condition to second order is 
\begin{align}
\begin{split}
	\frac{\sqrt{m}}{\sqrt{\Omega}} \theta_+ \Big|_{r = \rho}&\sim- (1 - m^{AB} D_A \rho D_B \rho) (p_v + y_v) + D_A \rho y^A - \frac{1}{2} D_A\rho D_B \rho \Omega^{AB} p_h\\
	&+ \qty(2 r - \frac{1}{2} X^{AB}/r \partial_r (X_{AB}/r) - D_A D^A \rho + r^{-2} D_A(X^{AB} D_B \rho))\Big|_{r=\rho} = 0
\end{split}
\end{align}
Note that all the quantities on the right-hand side implicitly depend on $\rho$.
This relation has to be solved for the function $\rho$. 
We apply the following strategy:
We perturbatively expand $\rho$ as $\rho = \rho^{(0)} + \rho^{(1)} + \rho^{(2)}$
where $\rho^{(i)}$ is of $i$-th order in the perturbations.
To deal with the implicit dependence on $\rho$ we use a Taylor series around $\rho^{(0)}$.
For the Schwarzschild solution the apparent horizon is purely symmetrical and we assume the same for $\rho^{(0)}$.
We have
\begin{align}
\begin{aligned}
    &\Big[- p_v^{(0)} - p_v^{(2)} - \partial_r p_v^{(0)} (\rho^{(1)} + \rho^{(2)}) - \frac{1}{2}\partial_r^2 p_v^{(0)} (\rho^{(1)})^2 - y_v^{(1)} - y_v^{(2)} - \partial_r y_v^{(1)} \rho^{(1)}+ D_A \rho^{(1)} y^A_{(1)} \\
    &+ r^{-2} D^A \rho^{(1)} D_A \rho^{(1)} p_v^{(0)} - \frac{1}{2} D^A \rho^{(1)} D_A \rho^{(1)} p_h^{(0)} + 2 (\rho^{(0)} + \rho^{(1)} + \rho^{(2)}) - \frac{1}{2} X^{AB}/r \partial_r (X_{AB}/r)\\
    & - D_A D^A (\rho^{(1)} + \rho^{(2)}) + r^{-2} D_A (X^{AB} D_B \rho^{(1)})\Big]_{r=\rho^{(0)}}=0
\end{aligned}
\label{eq:TrappingConditionTaylor}
\end{align}
All metric components are evaluated at $r = \rho^{(0)}$ after taking the derivatives. 

We now evaluate the equation order by order. 
At zeroth order we have $p_v^{(0)} = 2 \sqrt{r r_s}$ and the equation reads
\begin{equation}
    - 2\sqrt{\rho^{(0)} r_s} + 2 \rho^{(0)}=0
\end{equation}
The solutions are $\rho^{(0)}=0$ and $\rho^{(0)} = r_s$.
The latter is the physical one.
We find agreement with the intuition that the apparent horizon is a sphere with the Schwarzschild radius.

The first order equation is
\begin{align}
    \rho^{(1)} - y_v^{(1)} - D_A D^A \rho^{(1)} = 0
\end{align}
For the solution we expand $\rho^{(1)}$ in terms of spherical harmonics $\rho^{(1)} = \sum_{l\geq1,m}\rho^{(1)}_{lm} L_{lm}$. Then, the solution of the first order equation is
\begin{align}
    \rho^{(1)}_{lm} = \frac{y_{v,lm}^{(1)}}{l^2 + l + 1}\,.
\end{align}

Finally, we turn to the second order equation.  It is given by 
\begin{align}
    &- p_v^{(2)} - \rho^{(2)} + \frac{1}{4 r_s}(\rho^{(1)})^2 - y_v^{(2)} - \partial_r y_v^{(1)} \rho^{(1)} + r_s^{-2} D^A \rho^{(1)} D_A \rho^{(1)}p_v^{(0)} + D_A \rho^{(1)}y^A_{(1)} - \frac{1}{2} D^A \rho^{(1)} D_A \rho^{(1)} p_h^{(0)}\nonumber\\
    &+ 2 \rho^{(2)} - \frac{1}{2r_s^2} X^{AB} \partial_r X_{AB} + \frac{1}{2 r_s^3} X^{AB} X_{AB} - D_A D^A \rho^{(2)} + r_s^{-2} D_A(X^{AB} D_B \rho^{(1)}) = 0
\end{align}
We expand $\rho^{(2)}$ into spherical harmonics and solve the above equation for the non-symmetric ($l>0$) and symmetric ($l=0$) contributions.
For the computation of the non-symmetric contributions, we have to solve integrals with three scalar, vector and tensor spherical harmonics.
In \cite{V}, it is explained, how these integrals can be evaluated in terms of Clebsch-Gordan coefficients $C^{l m s}_{l' m' s', l'' m'' s''}$.
The computation is based on spin-weighted spherical harmonics, an alternative basis for vector and tensor spherical harmonics (see also \cite{SPSH}). 
The result for $l>0$ is
\small
\begin{align}
	(l^2+l+1)\rho^{(2)}_{lm} &= (y_v^{(2)})_{lm} - \sum_{l' m', l'' m''}\Big[ \qty(\frac{1}{4 r_s} \rho^{(1)}_{l' m'} \rho^{(1)}_{l''m''} - \partial_r (y_v^{(1)})_{l' m'} \rho^{(1)}_{l'' m''})C^{l m 0}_{l' m' 0, l'' m''0} \nonumber\\
				 &+ \sqrt{l'(l'+1)}\Big(-\sqrt{l''(l''+1)} \rho^{(1)}_{l' m'} \rho^{(1)}_{l'' m''} \sigma_+ \big(r_s^{-2} p_v^{(0)} - \frac{1}{2}p_h^{(0)}\big) - \sigma_+ \rho^{(1)}_{l' m'} y^e_{l'' m''} - i \sigma_- \rho^{(1)}_{l' m'} y^o_{l'' m''}\nonumber\\
			&~~~~~~~~~~~~~~+ r_s^{-2} \sqrt{\frac{(l''+2)(l''-1)}{2}} (i \sigma_- X^o_{l''m''} + \sigma_+ X^e_{l'' m''}) \rho^{(1)}_{l' m'}\Big) C^{lm 0}_{l' m' 1, l'' m''-1}\\
			&+ \sum_{IJ}\Big( \frac{1}{2r_s^3}X^I_{l' m'} X^J_{l'' m''} q_{IJ} - \frac{1}{2r_s^2} X^I_{l'm'} \partial_r X^J_{l'' m''} q_{IJ}\nonumber\\
            &~~~~~~~~~~~~~~+ r_s^{-2} \sqrt{\frac{(l'+2)(l'+1)l'(l'-1)}{2}} (i \sigma_- X^o_{l'' m''} + \sigma_+ X^e_{l'' m''}) \rho^{(1)}_{l' m'}\Big) C^{lm0}_{l' m' 2, l'' m''-2}\nonumber\,,
\end{align}
\normalsize
The objects $\sigma_{\pm}$ are projectors onto the subspace where the sum $l+l'+l''$ is even / odd respectively. The quantity $q_{IJ}$ is defined by $q_{ee}= q_{oo} = \sigma_+$ and $q_{eo} = -q_{oe} = i \sigma_-$.

The spherical symmetric contribution ($l=0$) to $\rho^{(2)}$ is 
\begin{align}
\begin{aligned}
    \rho_{00}^{(2)} &= \sqrt{4 \pi} p_v^{(2)} + \frac{1}{\sqrt{4\pi}}\Big[ \frac{y_v^{(1)} \cdot \partial_r y_v^{(1)}}{(l^2+l+1)} - \frac{(6 l(l+1)+1)y_v^{(1)}\cdot y_v^{(1)}}{4r_s(l^2+l+1)^2}\\
    &- \frac{\sqrt{l(l+1)}}{(l^2+l+1) } y_v^{(1)} \cdot y_e^{(1)} + \frac{1}{2r_s^2} (X^o \cdot \partial_r X^o + X^e \cdot \partial_r X^e) - \frac{1}{2r_s^3} (X^o \cdot X^o + X^e \cdot X^e)\Big]
\end{aligned}
\end{align}
In the above solutions we have to replace $p_v^{(2)}$, $y_v^{(1)}$ and $y_e^{(1)}$ in terms of the solution we found in the previous papers.
For two functions $f,g$ with expansion coefficients $f_{lm},g_{lm}$ we abbreviated
\begin{align}
	(f \cdot g) := \sum_{lm} f_{lm} g_{lm}
\end{align}
In summary, we now have full control over the shape of the apparent horizon to second order in the perturbations.

For the physics of black holes, the area of the apparent horizon is of particular interest.
The square root of this area defines a dynamical and quasi local notion of mass, called the \emph{quasi-local mass}. 
The area of the apparent horizon is computed by integrating the area element arising from the induced metric $h_{AB}$ over the apparent horizon.
We pull-back the metric $m_{ij}$ to the apparent horizon:
\begin{align}
    h_{AB} &= Y_{(A)}^i Y_{(B)}^j m_{ij} = \rho^2 \Omega_{AB} + X_{AB} + D_A \rho D_B \rho \\
    &\sim(r_s^2 + 2 r_s \rho^{(1)} + 2 r_s \rho^{(2)} + (\rho^{(1)})^2) \Omega_{AB} + X_{AB} + \partial_r X_{AB} \rho^{(1)} + D_A \rho^{(1)} D_B \rho^{(1)}\,.
\end{align}
where we approximated the induced metric to second order. $X_{AB}$ was expanded linearly using a Taylor approximation. 
In the calculation we already inserted the value of $\rho^{(0)}$ in terms of the Schwarzschild mass. 
The square root of the determinant of $h_{AB}$ to second order is given by
\begin{align}
    \sqrt{h} \sim \sqrt{\Omega}\qty[r_s^2 + 2 r_s \rho^{(1)} + 2 r_s \rho^{(2)} + (\rho^{(1)})^2 + \frac{1}{2} D_A \rho^{(1)} D^A \rho^{(1)} - \frac{1}{4 r_s^2}X_{AB} X^{AB}]
\end{align}
The area is then obtained by integrating the area element $\sqrt{h}$ over the sphere. We have
\begin{align}
    A = 4 \pi r_s^2 + 2 r_s \sqrt{4\pi} \rho^{(2)}_{00} + \frac{(2 + l(l+1)) y_v^{(1)}\cdot y_v^{(1)}}{2(l^2+l+1)^2} - \frac{1}{4 r_s^2}(X^o \cdot X^o + X^e \cdot X^e)
\end{align}
We now insert $\rho^{(2)}_{00}$ and obtain
\begin{align}
\begin{aligned}
    A &= 4 \pi r_s^2 + 8 \pi r_s p_v^{(2)}+ 2 r_s \Big[ \frac{y_v^{(1)} \cdot \partial_r y_v^{(1)}}{(l^2+l+1)} - \frac{\sqrt{l(l+1)}}{(l^2+l+1) } y_v^{(1)} \cdot y^e_{(1)}\Big]\\
    &+ \frac{(1 - 5 l(l+1)) y_v^{(1)}\cdot y_v^{(1)}}{2(l^2+l+1)^2} + \frac{1}{r_s}(X^o \partial_r X^o + X^e \partial_r X^e) - \frac{5}{4 r_s^2}(X^o \cdot X^o + X^e \cdot X^e)
\end{aligned}
\end{align}
Recall the relation of the momentum $ y_e^{(1)}$ with $ y_v^{(1)}$ and $ X^e$:
\begin{equation}
	\sqrt{l(l+1)}  y_e^{(1)} = 2 \partial_r   y_v^{(1)} - \frac{1}{r_s}  y_v^{(1)} + \frac{1}{2 r_s^2}\sqrt{2(l+2)(l+1)l(l-1)}  X^e
\end{equation}
Then, the area becomes
\begin{align}
\begin{aligned}
	A &= 4 \pi r_s^2 + 8 \pi r_s p_v^{(2)} -  2 r_s \frac{y_v^{(1)} \cdot \partial_r y_v^{(1)}}{(l^2+l+1)} - \frac{\sqrt{2(l+2)(l+1)l(l-1)}}{(l^2+l+1) r_s} y_v^{(1)} \cdot X^e\\
    &+ \frac{(5 - l(l+1)) y_v^{(1)}\cdot y_v^{(1)}}{2(l^2+l+1)^2} + \frac{1}{r_s}(X^o \partial_r X^o + X^e \partial_r X^e) - \frac{5}{4 r_s^2}(X^o \cdot X^o + X^e \cdot X^e)
\end{aligned}
\end{align}
Next, we introduce the second order solution for the background momentum. In Section \ref{sec:RevPert}  we found
\begin{align}
    p_v^{(2)}(r_s) = \frac{1}{8\pi} \qty[\int^{r_s} E \dd{r} + B_o(r_s) + B_e(r_s)]\,,
\end{align}
where $B_o$ and $B_e$ are the boundary terms of the odd and even polarity perturbations. 
The evaluation of the boundary terms for the Schwarzschild radius gives
\begin{align}
    B_o(r_s) &= - \frac{ X^o \cdot \partial_r  X^o}{r_s^2} + \frac{5}{4 r_s^3}  X^o \cdot  X^o\\
    \begin{split}
    B_e(r_s) &= - \frac{ X^e \cdot \partial_r  X^e}{r_s^2} + \frac{5}{4 r_s^3}  X^e \cdot  X^e + \frac{\sqrt{2(l+2)(l+1)l(l-1)}}{(l^2+l+1)r_s^2} X^e \cdot  y^{(1)}_v\\
	     &+ \frac{l^2+l-5}{2(l^2+l+1)^2 r_s} y_v^{(1)} \cdot  y_v^{(1)} + \frac{2}{l^2+l+1}  y_v^{(1)} \cdot \partial_r  y^{(1)}_v
     \end{split}
\end{align}
Replacing $p_v^{(2)}$, we find that many terms cancel. We are left with
\begin{align}
\begin{aligned}
	A &= 4 \pi r_s^2 + r_s \int^{r_s} E \dd{r}\,.
\end{aligned}
\end{align}

At zeroth order the apparent horizon is a round sphere of coordinate radius given by Schwarzschild radius
and its area is simply the area of that sphere with respect to the unperturbed 
Schwarzschild metric. At second order, the apparent horizon is still homeomorphic 
to a sphere but no longer one of constant coordinate radius as 
displayed explicitly by the angular dependent first and second order corrections to the radial profile 
computed above. However, the area of the apparent horizon is an integral over the angular 
coordinates and therefore itself not angle dependent. Rather, deviations of the shape of the apparent 
horizon from a round sphere express themselves by contributions from the perturbations 
to its area. The corresponding   
correction term to second order turns out to be given by an integral over the mass density $E(r)$
within a round sphere of radius $r_s$.

We define the quasi-local mass as $M_0:= \sqrt{A/(16\pi)}$ and have
\begin{align}
	M_0 &= M + \frac{1}{16 \pi} \int^{r_s} E \dd r
\end{align}
The result for $M_0$ is again surprisingly simple.
It only depends on the initial mass $M$ and the integral over the local mass density inside a round sphere 
of Schwarzschild radius.

\subsection{Extension to electromagnetic fields}

In this section, we generalise the analysis of the apparent horizon from pure gravity to its coupling with electromagnetic matter.
The Hamiltonian formulation of the Einstein-Maxwell system in GP gauge to second order was developed in \cite{III}. 
In this section we briefly recall some formulas necessary for the computation and refer the reader to \cite{III} for more details. 
Then, we discuss the solution of the trapping condition and determine the area of the apparent horizon and the quasi-local mass.

We use the same definition for the gravitational variables as before.
In addition, we introduce the vector potential and electric field $(A_i,E^i)$ and decompose them into spherical harmonics:
\begin{align}
    A_3 &= \sum_{l\geq 1, m} x_M^{lm}L_{lm}, \quad E^3 = \sqrt{\Omega} \qty(\xi + \sum_{l\geq 1, m} y^M_{lm} L_{lm})\\
    A_B &= \sum_{l\geq 1, m, I} X^I_{lm} [L_{I,lm}]_B, \quad E^B = \sqrt{\Omega} \sum_{l \geq 1, m, I} Y^M_{I,lm} [L_{I,lm}]^B
\end{align}
The Gauss constraint implies that $\xi$ is a constant and it is related to the electric charge of the black hole in a Reissner-Nordstr\o{}m geometry.
We gauge fix the electromagnetic gauge freedom by setting $x_M^{lm} =0$ and we obtain the solution of the Gauss constraint 
\begin{equation}
    y_{lm}^M = \sqrt{l(l+1)} \int Y^M_{e,lm} \dd r
\end{equation}

As in section \ref{sec:RevPert}, we now expand the Hamiltonian and diffeomorphism constraints in terms of the non-symmetric degrees of freedom $(l>0)$ and 
solve them order by order.
The details of this computation can be found in \cite{III}.
Here, we only summarise the results.

At zeroth order we find the solutions
\begin{align}
    p_v^{(0)} &= 2 \sqrt{r r_s - \frac{g^2 \xi^2}{4}}\\
    p_h^{(0)} &= \frac{2 r_s}{p_v^{(0)} r}
\end{align}
This describes the Reissner-Nordstr\o{}m black hole in GP coordinates with electric charge $Q = g \xi$. 

At first order, for the odd polarity constraints, we obtain
\begin{equation}
    (y_o^{(1)})_{lm} = \frac{1}{r^2}\int \qty[\sqrt{2(l+2)(l+1)l(l-1)}\qty(r^2 Y_o^{lm} + \frac{p_h^{(0)}}{2} X^o_{lm}) - \xi \partial_r X^{o,lm}_M]\dd r
\end{equation}
For the even polarity sector, we have
\begin{align}
    \begin{aligned}
    (y_h^{(1)})_{lm} &= \frac{1}{r^2}\qty(1 - \frac{r^2 p_h^{(0)}}{p_v^{(0)}}) [y_v^{(1)}]_{lm} - \frac{\sqrt{2(l+2)(l+1)l(l-1)}}{2 r^2 p_v^{(0)}} X^e_{lm} + \frac{g^2 \xi \sqrt{l(l+1)}}{r p_v^{(0)}} \int Y^M_{e,lm} \dd r \\
    \sqrt{l(l+1)}& [y_e^{(1)}]_{lm} = - 2 \partial_r [y_h^{(1)}]_{lm} + 2 \partial_r [y_v^{(1)}]_{lm}
    \end{aligned}
\end{align}
In addition, $y_v$ has to satisfy a differential equation similar to \eqref{eq:yv} which we do not repeat here and can be found in \cite{III}.

The solution of the second order constraints can be put into the following form:
\begin{align}
    p_v^{(2)} &= \frac{r}{4 \pi p_v^{(0)}} \qty[\int E \dd r + B_o + B_e]\\
    p_h^{(2)} &= \frac{1}{r^2}\qty(1- \frac{r^2 p_h^{(0)}}{p_v^{(0)}}) p_v^{(2)} + \frac{1}{4 \pi p_v^{(0)}} {}^{(2)}C_v
\end{align}
where $B_o$ and $B_e$ are boundary terms depending on the odd and even polarity perturbations respectively.
The odd polarity boundary term is
\begin{align}
    B_o = \frac{r - r_s + \frac{g^2 \xi^2}{4r}}{r^4}\qty[\int \qty(r^2 Y_o + \frac{p_v}{2}) X^o \dd r]^2 - \frac{1}{r^2} X^o \cdot \partial_r X^o - \frac{6 r - r_s}{4 r^4} X^o \cdot X^o
    \label{eq:BoEM}
\end{align}
The even polarity boundary term is recorded in appendix \ref{sec:EvenParityBdryEM}.

Then, in \cite{III}, we find the reduced Hamiltonian for one asymptotic end:
\begin{equation}
    H = \lim_{r\to \infty} \frac{2\pi}{\kappa r} p_v^2 = M + \frac{1}{\kappa}\int_{\mathbb{R}^+} E \dd r
\end{equation}
$E$ has the interpretation as a local energy density and can be written in terms of master variables.
In \cite{III} it is shown that the boundary terms vanish at infinity, so in the limit $r\to \infty$ they drop out of the physical Hamiltonian.
The density $E$ is of the form
\begin{equation}
    E = \sum_I \qty[\frac{p_v^{(0)}}{2r} P_I \partial_r Q_I - \frac{1}{2}\qty(P_I^2 + (Q_I')^2 + V_I Q_I^2)]
\end{equation}
where $(Q_I,P_I)$ are the master variables, $V_I$ are the Regge-Wheeler-Zerilli potentials.
The index $I$ runs over the modes of the master variable (even/odd and electromagnetic/gravity).\\

After this brief review, we turn to the analysis of the apparent horizon.
In the Gullstrand-Painleve gauge, the components $X^e$ and $X^o$ of the metric are non-vanishing.
The electromagnetic degrees of freedom enter the equations defining the apparent horizon through the solution of the constraint equations for the gauge momenta $p,y$. 
For this reason, we can reuse a lot of the analysis of the previous section.
In particular we have to solve equation \eqref{eq:TrappingConditionTaylor} to determine the function $\rho(\theta,\phi)$.

The zeroth order equation now reads using the solution for $p_v^{(0)}$
\begin{equation}
    2 \rho^{(0)} - p_v^{(0)} = 2 \rho^{(0)} - 2 \sqrt{r r_s - \frac{1}{4} g^2 \xi^2} = 0
\end{equation}
As before, the equation has two roots which we denote by $r_\pm$.
In contrast to the purely gravitational setup, the roots correspond to the inner and outer horizon of the Reissner-Nordstr\o{}m black hole.
Both of them are physically significant and we will perform the analysis for both of them simultaneously.
The solutions are given by
\begin{equation}
    r_{\pm} = \frac{1}{2}\qty(r_s \pm \sqrt{r_s^2 - g^2 \xi^2}) 
\end{equation}
Form this equation it is easy to see that $r_+ + r_-=r_s$ and $r_+ r_- = g^2 \xi^2/4$.
Notice that by the zeroth order equation, we have:
\begin{equation}
    p_v^{(0)} \Big|_{r = r_\pm} = 2 r_\pm
\end{equation}
This identity will be useful for the following computations.

At first order, we obtain
\begin{equation}
    \qty[\qty(l(l+1) + 2 - \partial_r p_v^{(0)} ) \rho^{(1)}_{lm} - y_{v,lm}^{(1)}]_{r = r_\pm} = 0
\end{equation}
Comparing with the purely gravitational case, we see that the difference is the value of $\partial_r p_v^{(0)}$ evaluated at $r_\pm$.
Before, it was equal to $1$, but now we have using $\partial_r p_v^{(0)} = 2 r_s / p_v^{(0)}$
\begin{equation}
    \qty(l(l+1) + 2 - \frac{r_s}{r_\pm} ) \rho^{(1)}_{lm} - y_{v,lm}^{(1)} (r_\pm) = 0
\end{equation}
Thus, the solution of the first-order equation for $\rho_{lm}^{(1)}$ is
\begin{equation}
    \rho^{(1)}_{lm} = \frac{y_{v,lm}^{(1)}(r_\pm)}{l(l+1) + 2 - r_s/r_\pm}
\end{equation}

At second order we have the equation
\begin{align}
\begin{aligned}
    &\Big[\sum_{lm} (l(l+1) + 2 - \partial_r p_v^{(0)}) \rho^{(2)}_{lm} L_{lm} - p_v^{(2)} - \frac{1}{2}\partial_r^2 p_v^{(0)} (\rho^{(1)})^2 - y_v^{(2)} - \partial_r y_v^{(1)} \rho^{(1)} + r^{-2} D^A \rho^{(1)} D_A \rho^{(1)} p_v^{(0)}\\
    &+ D_A \rho^{(1)} y^A_{(1)} - \frac{1}{2} D^A \rho^{(1)} D_A \rho^{(1)} p_h^{(0)} - \frac{1}{2} X^{AB}/r \partial_r (X_{AB}/r) + r^{-2} D_A (X^{AB} D_B \rho^{(1)})\Big]\Big|_{r=\rho^{(0)}}=0
\end{aligned}
\end{align}
For the computation, we need to evaluate the second derivative of $p_v^{(0)}$ and the value $p_h^{(0)}(r_\pm)$.
We have
\begin{align}
    \partial_r^2 p_v^{(0)} &= - \frac{4 r_s^2}{(p_v^{(0)})^3}\Bigg|_{r=r_\pm}  = - \frac{r_s^2}{2 r_\pm^3}\\
    p_h^{(0)}(r_\pm) &= \frac{2 r_s}{p_v^{(0)} r}\Bigg|_{r=r_\pm} = \frac{r_s}{r_\pm^2} 
\end{align}

For the non-symmetric contributions to the equation, with $l>0$, we have
\small
\begin{align}
	\qty(l(l+1) + 2 - \frac{r_s}{r_\pm})\rho^{(2)}_{lm} &= (y_v^{(2)})_{lm} - \sum_{l' m', l'' m''}\Big[ \qty(\frac{r_s^2}{4 r_\pm^3} \rho^{(1)}_{l' m'} \rho^{(1)}_{l''m''} - \partial_r (y_v^{(1)})_{l' m'} \rho^{(1)}_{l'' m''})C^{l m 0}_{l' m' 0, l'' m''0} \nonumber\\
	&+ \sqrt{l'(l'+1)}\Big(\frac{r_s - 4 r_\pm}{2 r_\pm^2}\sqrt{l''(l''+1)} \rho^{(1)}_{l' m'} \rho^{(1)}_{l'' m''} \sigma_+ - \sigma_+ \rho^{(1)}_{l' m'} y^e_{l'' m''} - i \sigma_- \rho^{(1)}_{l' m'} y^o_{l'' m''}\nonumber\\
	&~~~~~~~~~~~~~~+ r_\pm^{-2} \sqrt{\frac{(l''+2)(l''-1)}{2}} (i \sigma_- X^o_{l''m''} + \sigma_+ X^e_{l'' m''}) \rho^{(1)}_{l' m'}\Big) C^{lm 0}_{l' m' 1, l'' m''-1}\\
	&+ \sum_{IJ}\Big( \frac{1}{2r_\pm^3}X^I_{l' m'} X^J_{l'' m''} q_{IJ} - \frac{1}{2r_\pm^2} X^I_{l'm'} \partial_r X^J_{l'' m''} q_{IJ}\nonumber\\
    &~~~~~~~~~~~~~~+ r_\pm^{-2} \sqrt{\frac{(l'+2)(l'+1)l'(l'-1)}{2}} (i \sigma_- X^o_{l'' m''} + \sigma_+ X^e_{l'' m''}) \rho^{(1)}_{l' m'}\Big) C^{lm0}_{l' m' 2, l'' m''-2}\nonumber\,,
\end{align}
\normalsize

On the other hand, for the symmetric contribution, we find
\begin{align}
    \begin{aligned}
        \qty(2 - \frac{r_s}{r_\pm}) &\rho^{(2)}_{lm} = \sqrt{4\pi} p_v^{(2)} + \frac{1}{\sqrt{4\pi}}\Big[\frac{1}{2 r_\pm^2} (X^e \cdot \partial_r X^e + X^o \cdot \partial_r X^o) - \frac{1}{2 r_\pm^3} (X^e \cdot X^e + X^o \cdot X^o) \\
        &+ \rho^{(1)} \cdot \partial_r y_v^{(1)} - \sqrt{l(l+1)} \rho^{(1)} \cdot y_e^{(1)} + \frac{- r_s^2 + 2 l(l+1) r_s r_\pm - 8 l(l+1) r_\pm^2}{4 r_\pm^3}\rho^{(1)} \cdot \rho^{(1)} \Big]_{r=\rho^{(0)}}
    \end{aligned}
\end{align}
Inserting the solution of the first order equation, we obtain
\begin{align}
        \qty(2 - \frac{r_s}{r_\pm}) &\rho^{(2)}_{00} = \sqrt{4\pi} p_v^{(2)}(r_\pm) + \frac{1}{\sqrt{4\pi}}\Big[\frac{1}{2 r_\pm^3} \sum_{I \in \{e,o\}}\qty( r_\pm^2 X^I \cdot \partial_r X^I - X^I \cdot X^I) + \frac{1}{l(l+1) + 2 - \frac{r_s}{r_\pm}} y_v^{(1)} \cdot \partial_r y_v^{(1)}\nonumber\\
        &- \frac{\sqrt{l(l+1)}}{l(l+1) + 2 - \frac{r_s}{r_\pm}} y_v^{(1)} \cdot y_e^{(1)} + \frac{- r_s^2 + 2 l(l+1) r_s r_\pm - 8 l(l+1) r_\pm^2}{4 r_\pm^3(l(l+1) + 2 - \frac{r_s}{r_\pm})^2}y_v^{(1)} \cdot y_v^{(1)} \Big]_{r=\rho^{(0)}}
\end{align}

For the area element on the horizon ($r=r_\pm$), we have using a similar computation as before
\begin{equation}
    \sqrt{h} \sim \sqrt{\Omega}\qty[r_\pm^2 + 2 r_\pm \rho^{(1)} + 2 r_\pm \rho^{(2)} + (\rho^{(1)})^2 + \frac{1}{2} D_A \rho^{(1)} D^A \rho^{(1)} - \frac{1}{4 r_\pm^2}X_{AB} X^{AB}]
\end{equation}
In order to obtain the area, we have to integrate the area element over the sphere with radius $r_\pm$. 
Performing the integral, we find
\begin{equation}
    A = 4 \pi r_\pm^2 + 2 r_\pm \sqrt{4\pi} \rho^{(2)}_{00} + \frac{(2 + l(l+1)) y_v^{(1)}\cdot y_v^{(1)}}{2(l(l+1) + 2 - \frac{r_s}{r_\pm})^2} - \frac{1}{4 r_\pm^2}(X^o \cdot X^o + X^e \cdot X^e)
\end{equation}
As before, the area consists of a zeroth order contribution and a quadratic correction. 
For its final form, we insert $\rho^{(2)}_{00}$ and we obtain
\begin{align}
    \begin{aligned}
    A &= 4 \pi r_\pm^2 + 8 \pi r_\pm p_v^{(2)} + \frac{2 r_\pm}{2 - \frac{r_s}{r_\pm}}\Big[\frac{1}{2 r_\pm^2} (X^e \cdot \partial_r X^e + X^o \cdot \partial_r X^o) - \frac{6 - r_s/r_\pm}{8 r_\pm^3} (X^e \cdot X^e + X^o \cdot X^o) \\
    &+ \frac{1}{l(l+1) + 2 - \frac{r_s}{r_\pm}} y_v^{(1)} \cdot \partial_r y_v^{(1)} - \frac{\sqrt{l(l+1)}}{l(l+1) + 2 - \frac{r_s}{r_\pm}} y_v^{(1)} \cdot y_e^{(1)}\\
    &+ \frac{- r_s^2 + (l+2)(l-1) r_s r_\pm + (4 - 6 l(l+1)) r_\pm^2}{4 r_\pm^3(l(l+1) + 2 - \frac{r_s}{r_\pm})^2}y_v^{(1)} \cdot y_v^{(1)} \Big]_{r=\rho^{(0)}}
    \end{aligned}
\end{align}

Next, we need to use the solution for the momentum $y_e^{(1)}$ in terms of $y_v^{(1)}$, $X^e$ and $A^e$.
The relation is then evaluated on the horizon:
\begin{align}
    \sqrt{l(l+1)}y_e^{(1)} &= 2 \partial_r y_v^{(1)} -\frac{2}{r}\qty(1 - \frac{r^2 p_h^{(0)}}{p_v^{(0)}}) y_v^{(1)} + \frac{\sqrt{2(l+2)(l+1)l(l-1)}}{r p_v^{(0)}} X^e + \frac{2 g^2 \xi \sqrt{l(l+1)}}{r p_v^{(0)}} A^e\\
    \sqrt{l(l+1)}y_e^{(1)}(r_\pm) &= 2 \partial_r y_v^{(1)}(r_\pm) -\frac{2 r_\pm - r_s}{r_\pm^2} y_v^{(1)} + \frac{\sqrt{2(l+2)(l+1)l(l-1)}}{2 r_\pm^2} X^e + \frac{g^2 \xi \sqrt{l(l+1)}}{r_\pm^2} A^e
\end{align}
We find the solution for the area
\begin{align}
    \begin{aligned}
    A &= 4 \pi r_\pm^2 + 8 \pi r_\pm p_v^{(2)} + \frac{2 r_\pm}{2 - \frac{r_s}{r_\pm}}\Big[\frac{1}{2 r_\pm^2} (X^e \cdot \partial_r X^e + X^o \cdot \partial_r X^o) - \frac{6 - r_s/r_\pm}{8 r_\pm^3} (X^e \cdot X^e + X^o \cdot X^o) \\
    &- \frac{1}{l(l+1) + 2 - \frac{r_s}{r_\pm}} y_v^{(1)} \cdot \partial_r y_v^{(1)}  - \frac{\sqrt{2(l+2)(l+1)l(l-1)}}{2 r_\pm^2 (l(l+1) + 2 - \frac{r_s}{r_\pm})} y_v^{(1)} \cdot X^e\\
    & - \frac{g^2 \xi \sqrt{l(l+1)}}{r_\pm^2 (l(l+1) + 2 - \frac{r_s}{r_\pm})} y_v^{(1)} \cdot A^e+ \frac{3 r_s^2 - 3 (6 + l(l+1)) r_s r_\pm + 2(10 + l(l+1)) r_\pm^2}{4 r_\pm^3(l(l+1) + 2 - \frac{r_s}{r_\pm})^2}y_v^{(1)} \cdot y_v^{(1)} \Big]_{r=\rho^{(0)}}
    \end{aligned}
\end{align}

In the last step, we have to replace $p_v^{(2)}$ by its solution in terms of an integral over the Hamiltonian density and the boundary terms. 
For this recall the boundary terms \eqref{eq:BoEM} and \eqref{eq:BeEM} and evaluate them for $r = r_\pm$:
\begin{align}
    B_o = - \frac{1}{r_\pm^2} X^o \cdot \partial_r X^o + \frac{6 r_\pm - r_s}{4 r_\pm^4} X^o \cdot X^o
\end{align}
Since the even polarity boundary term is more complicated, we use Mathematica and obtain
\begin{align}
    B_e &= - \frac{1}{r_\pm^2} X^e \cdot \partial_r X^e+ \frac{6 r_\pm - r_s}{4 r_\pm^4} X^e \cdot X^e + \frac{2}{l(l+1) + 2 - \frac{r_s}{r_\pm}} y_v^{(1)} \cdot \partial_r y_v^{(1)} + \frac{2 g \xi \sqrt{l(l+1)}}{r_\pm^2 (l(l+1) + 2 - \frac{r_s}{r_\pm})} y_v^{(1)} \cdot A^e\nonumber\\
    &+ \frac{\sqrt{2(l+2)(l+1)l(l-1)}}{(l(l+1) + 2 - \frac{r_s}{r_\pm}) r_\pm^2} y_v^{(1)} \cdot X^e  + \frac{-3 r_s^2 + 3 (l^2+l+6) r_s r_\pm -2(l^2+l+10)r_\pm^2}{2 r_\pm^3(l(l+1) +2 - \frac{r_s}{r_\pm})^2} y_v^{(1)} \cdot y_v^{(1)}
\end{align}
In terms of $B_o$ and $B_e$, the expression for the area of the apparent horizon simplifies drastically:
\begin{equation}
    A = 4 \pi r_\pm^2 + 8 \pi r_\pm p_v^{(2)}(r_\pm) - \frac{r_\pm}{2 - \frac{r_s}{r_\pm}}\qty(B_o + B_e)
\end{equation}
On the horizon, the solution of the second order constraints $p_v^{(2)}$ simplifies and we have
\begin{equation}
    p_v^{(2)}(r_\pm) = \frac{1}{8 \pi} \int^{r_\pm} E \dd r + B_o(r_\pm) + B_e(r_\pm)
\end{equation}
Inserting this into the formula for the area of the apparent horizon, we find it in its final form:
\begin{equation}
    A = 4 \pi r_\pm^2 + r_\pm \int^{r_\pm} E \dd r + \frac{r_\pm - r_s}{2 - \frac{r_s}{r_\pm}}\qty(B_o + B_e)
\end{equation}
Finally, the quasi-local mass is given to second order by 
\begin{equation}
    M_0 = \sqrt{\frac{A}{16 \pi}} \approx \frac{1}{2} r_\pm + \frac{1}{16 \pi} \qty[\int^{r_\pm} E \dd r + \frac{r_\pm - r_s}{2 r_\pm - r_s}\qty(B_o + B_e)]
\end{equation}
Notice that the fraction in front of the boundary terms can also be written as
\begin{equation}
    \frac{r_\pm - r_s}{2 r_\pm - r_s} = \frac{r_\pm - (r_+ + r_-)}{2 r_\pm - (r_+ + r_-)} = - \frac{r_\mp}{r_\pm - r_\mp}
\end{equation}
For the case that $r_-\to 0$ and $r_+ \to r_s$ (zero charge limit), the boundary terms vanish and we recover the previous result.

\subsection{Time-dependence of the Quasi-local Mass and Energy Flux through the Horizon}

To evaluate the time dependence of the quasi-local mass $M_0$, we need to calculate its Poisson bracket with the physical Hamiltonian. 
In the review section, we saw that the physical Hamiltonian is given by an integral over a quantity $E$ which has the interpretation of an energy density.

For the computation, we start with a generic form for $E$:
\begin{equation}
    E = N^3 P Q' + \frac{N}{2} (P^2 + (Q')^2 + V Q^2)
\end{equation}
where $(Q,P)$ is a canonical pair, $V$ a potential (mass term) and $N,N^3$ the lapse and radial shift. 
Notice, both the purely gravitational and the electromagnetic case in the GP gauge are given by the choice $N = 1$ and $N^3 = p_v^{(0)}/(2r)$.
Defining the smeared density by $E[f]:=\int \dd r E(r) f(r)$, we compute the Poisson bracket
\begin{equation}
    \{E[f],E[g]\}= \int \dd r (f g' - g f') \qty(N N^3 (P^2 + (Q')^2) + (N^2 + (N^3)^2) P Q')
    \label{eq:PBEEGeneric}
\end{equation}
The physical Hamiltonian is given by the constant smearing function $f=1$ and the local energy density $E(r_0)$ is obtained by setting $f=\delta(r-r_0)$.

The first application of the Poisson bracket \eqref{eq:PBEEGeneric} is the definition of a conserved current. 
We define the energy density $j^0$ as the local energy density $E$ and would like to determine the flux $j^3$. 
For the computation, we choose $g=1$ and $f=\delta(r-r_0)$ in \eqref{eq:PBEEGeneric} and obtain
\begin{equation}
    \partial_t j^0(r_0) = \{E(r_0),H_{\mathrm{phys}}\} = \partial_r\qty[N N^3 (P^2 + (Q')^2) + (N^2 + (N^3)^2) P Q']_{r = r_0}
\end{equation}
Thus, we obtain the continuity equation $\partial_t j^0 = -\partial_3 j^3$ with
\begin{align}
    j^0 &= N^3 P Q' + \frac{N}{2} (P^2 + (Q')^2 + V Q^2)\\
    j^3 &= - \qty[N N^3 (P^2 + (Q')^2) + (N^2 + (N^3)^2) P Q']
\end{align}
In order to get more intuition, we use the fact that from the equation of motion for $\dot Q$, we have
\begin{equation}
    P = \frac{1}{N}(\dot Q - N^3 Q')
\end{equation}
Then, we obtain
\begin{align}
    \begin{split}
    j^0 &= \frac{1}{2N}\qty[(\dot Q)^2 + (N^2 - (N^3)^2)(Q')^2 + N^2 V Q^2]\\
    &=\frac{1}{2}\sqrt{-g}\qty[g^{tt} (\dot Q)^2 + g^{33} (Q')^2 + V Q^2]
    \end{split}\\
    \begin{split}
    j^3 &= - \frac{1}{N} \dot Q\qty(N^3 \dot Q + (N^2 - (N^3)^2) Q')\\
    &= - \sqrt{-g} g^{3 \mu} Q_{,t} Q_{,\mu}
    \end{split}
\end{align}
where we used that $g^{tt}=-1/N^2$, $g^{3t}=N^3/N^2$, $g^{33}=(N^2-(N^3)^2)/N^2$ and $\sqrt{-g}=N$.
The above equation is consistent with what Moncrief found in equation (5.35) of \cite{6} for the gravitational case and in Schwarzschild coordinates.
To compare, we need to set $\sqrt{-g}=1$ and $g^{33}=1 - \frac{r_s}{r}$, $g^{3t}=0$ in the equation for $j^3$.
Notice that in Moncrief's equation there is a typo and the corrected relation in the paper should read
\begin{equation}
    \mathcal{\mathcal{H}^*}_{,0} = \partial_r\qty(\qty(1-\frac{r_s}{r}) Q_{,t} Q_{,r}) = \partial_r\qty(Q_{,t} Q_{,r_*})
\end{equation}
where $r_*$ is the turtoise coordinate defined by $\dv{r}{r_*}= 1- r_s/r$. 
Similarly in (6.17) of \cite{Moncrief1975}, the equation is generalised to the Reissner-Nordstr\o{}m black hole in Schwarzschild like coordinates.
Both of them are consistent with the continuity equation we derived in the Gullstrand Painlevé gauge above.

Let us now specialise to the purely gravitational case and study the time dependence of the quasi-local mass $M_0$. 
The quasi-local mass is given for the choice of smearing function $\theta(r_s -r)$, where $\theta(x)$ is the Heaviside theta function defined as
\begin{align}
	\theta(x) = \begin{cases}
		1 & \text{if } x >1\,,\\
		0 & \text{otherwise}
	\end{cases}
\end{align}
The physical Hamiltonian is given by the constant smearing function 1. 
We find using the distributional identity $\theta'(x) = \delta(x)$:
\begin{align}
	\dot M_{0} &= \qty{M_{0},H_\mathrm{phys}} = \sum_I \sqrt{\frac{r_s}{r}}\qty(P_I \cdot P_I + \partial_r Q^I \cdot \partial_r Q^I) + \qty(1 + \frac{r_s}{r}) P_I \cdot \partial_r Q^I\Big|_{r = r_s}\\
    &=\sum_I \qty(P_I + \partial_r Q^I) \cdot \qty(P_I + \partial_r Q^I)\Big|_{r=r_s}
\end{align}
In the equation we used that $N=1$ and $N^3 = \sqrt{r_s/r}$.
We observe that the change of the quasi-local mass is always positive.
Consistent with the black hole area law the quasi-local mass defined by the Gullstrand-Painleve observer 
is not allowed to decrease in classical general relativity.
A further simplification occurs using the equations of motion to substitute for the momentum $P$. 
We have
\begin{align}
    P_I = \dot Q^I - \sqrt{\frac{r_s}{r}}\partial_r Q^I
\end{align}
Then, the time derivative of the quasi-local mass is $\dot M_{0} = \sum_I \dot Q^I(r_s) \cdot \dot Q^I(r_s)$.\\
\\
We expect that in the quantum theory, when we replace classical objects in terms of annihilation and 
creation operators, then due to normal ordering effects there will be contributions to the quantum 
analog to $\dot{M}_0$ which are not manifestly positive operator valued. The direct computation of these
terms is beyond the scope of the present paper as it requires the detailed knowledge of the 
corresponding mode functions which will be the subject of another work in this series.

\section{Lapse and Shift in Terms of Reduced Phase Space Variables}
\label{sec:LapseShift}

In this section we derive the bulk solution for the perturbed lapse and shift perturbatively up to second order.
We focus on the vacuum case.
The radial shift is of special interest because in Gullstrand-Painleve coordinates 
it contains information about the local mass density as explained in \cite{I}.

We consider the GP gauge fixing condition and evaluate its stability under Hamiltonian evolution. 
This gives differential equations for lapse and shift with solutions $S = S_*$.
In \cite{I}, the exact, non-perturbative, asymptotic solutions were found. 
Here, we are interested in the bulk solution and use perturbation theory.
The computation requires the full Hamiltonian constraint $V_0$ and the full diffeomorphism constraint $V_\mu$ which are given by

\begin{gather}
    V_0 = \frac{1}{\sqrt{\det m}}\qty(m_{i k} m_{j l}- \frac{1}{2} m_{ij} m_{kl}) W^{ij} W^{kl} - \sqrt{\det m} R\\
    V_i = W^{jk} \partial_i m_{jk} - 2 \partial_j(W^{jk} m_{i k}) 
\end{gather}

We smear these constraints with suitable test functions $S^i_\parallel$, $S_\bot$.
$V_{\parallel}[S_\parallel]$ is defined by $\int V_i S^i_\parallel$ and $V_{\bot}[S_\bot]$ is given by $\int V_0 S_\bot$.
In \cite{I} we discussed that in order to make the variational principle well defined, we have to add suitable boundary terms.
The improved constraints are denoted by $H_\parallel[S_\parallel] := V_\parallel[S_\parallel] + B_\parallel[S_\parallel]$ and $H_\bot[S_\bot] := V_\bot[S_\bot] + B_\bot[S_\bot]$, where $B_\parallel, B_\bot$ are the suitably chosen boundary terms.

The GP gauge fixing conditions are $G_1 = m_{33}-1=0$, $G_2=m_{3A}=0$ and $G_3 = \Omega^{AB} m_{AB} - 2 r^2=0$.
They involve only the metric $m_{ij}$ and therefore we calculate: 
\begin{align}
    \qty{m_{ij}, H_\bot[S_\bot]} = \tilde S_\bot \qty(2 m_{i k} m_{jl} W^{kl} - m_{ij} m_{kl} W^{kl}),
\end{align}
where $\tilde S_\bot = S_\bot/\sqrt{m}$. The diffeomorphism constraint gives
\begin{align}
    \qty{m_{ij}, H_\parallel[S_\parallel]} = S^k_\parallel \partial_k m_{ij} + 2 m_{k (i} \partial_{j)} S^k_\parallel
\end{align}

We now specialise the variables to the GP gauge. The metric components are $m_{33}=1$, $m_{3 A}=0$ and $m_{AB} = r^2 \Omega_{AB} + X_{AB}$, where $X_{AB}$ is a symmetric, traceless tensor. 
The conjugate momenta are $W^{33}=\sqrt{\Omega}(p_v + y_v)$, $W^{3A} = \sqrt{\Omega} y^A/2$ and $W^{AB} = \sqrt{\Omega}(p_h + y_h)\Omega^{AB}/2 + \sqrt{\Omega} Y^{AB}$ as before.
For the test functions we use the convention $S^3_\parallel = f_h + g_h$ and $S^A_\parallel = g^A$ and $\tilde S_\bot = \sqrt{\Omega}^{-1}(f_v + g_v)$. 

The stability condition for the GP gauge ($G_1 = G_2 = G_3 = 0$) is given by the following Poisson brackets: 
\begin{align}
    \{G_1,H_\bot[S_\bot] +  H_\parallel[S_\parallel]\} &= 2 \partial_3 (f_h + g_h) +(f_v+g_v)((p_v + y_v) - r^2 (p_h + y_h) - X_{AB} Y^{AB}) = 0\nonumber\\
    \label{eq:GPfixingEqn1}
    \{G_2,H_\bot[S_\bot] +  H_\parallel[S_\parallel]\} &= \partial_A g_h + (r^2 \Omega_{AB} + X_{AB})\partial_3 g^B + (f_v+g_v) (r^2 \Omega_{AB} + X_{AB}) y^B =0\\
    \{G_3,H_\bot[S_\bot] +  H_\parallel[S_\parallel]\} &= 4 r (f_h + g_h) + 2 r^2 D_A g^A + 2 \Omega^{AB} X_{AC} D_B g^C + (f_v+g_v)\Big[(X^{AB}X_{AB}) (p_h + y_h)\nonumber\\
    &~~~~~~~~~~+ 2\Omega^{AB}X_{AC}X_{BD} Y^{CD} - 2 r^2 (p_v + y_v - X_{AB} Y^{AB})\Big]=0\nonumber
\end{align}
These are three coupled differential equations for the unknowns $f_h,g_h,f_v,g_v,g^A$.

\subsection{The zeroth order}
At zeroth order, we should recover the lapse and shift of the Schwarzschild metric expressed in Gullstrand-Painlevé coordinates.
We have the equations
\begin{align}
    2 \partial_3 f_h^{(0)} + f_v^{(0)}(p_v^{(0)} - r^2 p_h^{(0)}) &=0\\
    4 r f_h^{(0)} - 2 r^2 f_v^{(0)} p_v^{(0)} &=0
\end{align}
Solving the second equation for $f_v^{(0)}$ and inserting into the first gives
\begin{equation}
    \partial_3 f_h^{(0)} + \frac{1}{r p_v^{(0)}}(p_v^{(0)} - r^2 p_h^{(0)}) f_h^{(0)} = 0
\end{equation}
We know that $p_v^{(0)} = 2\sqrt{r r_s}$ and $p_h^{(0)} = r^{-2} \sqrt{r r_s}$. 
\begin{equation}
    \partial_3 f_h^{(0)} + \frac{1}{2r} f_h^{(0)} = 0
\end{equation}
Therefore $f_h^{(0)} = \sqrt{r_s/r}$, where we chose the integration constant to match the Schwarzschild metric in Gullstrand-Painlevé coordinates. 
The remaining zeroth order lapse is $f_v^{(0)} = r^{-2}$ and therefore $S_\bot = 1$.
For the full four-dimensional metric this means that we have $g_{tt} = - S_\bot + m_{33} (f_h^{(0)})^2 = - \qty(1 - \frac{r_s}{r})$ and $g_{tr} = m_{33} f^{(0)}_h = \sqrt{\flatfrac{r_s}{r}}$.
Thus, together with $m_{33} = 1$ and $m_{AB} = r^2 \Omega_{AB}$, we recover the Schwarzschild metric in Gullstrand-Painlevé coordinates as expected.

\subsection{The first order}
Next, the first order equations are given by
\begin{align}
    \begin{split}
    2 \partial_3 g_h^{(1)} + \frac{1}{r^2} (y_v^{(1)} - r^2 y_h^{(1)}) + g_v^{(1)}(p_v^{(0)} - r^2 p_h^{(0)})=0\\
    D_A g_h^{(1)} + r^2 \partial_3 g_A^{(1)} + y_A^{(1)} =0\\
    4 r g_h^{(1)} +2 r^2 D_A g^A_{(1)} - 2 y_v^{(1)}  - 2 r^2 g_v^{(1)} p_v^{(0)}= 0
    \end{split}
\end{align}
We first solve the last equation for $g_v^{(1)}$:
\begin{align}
    g_v^{(1)} = \frac{1}{2 r^2 p_v^{(0)}}\qty(4 r g_h^{(1)} + 2 r^2 D_A g^A_{(1)} - 2 y_v^{(1)})
\end{align}
Then, the first two equations are
\begin{align}
    \begin{split}
    4 r^2\partial_3 g_h^{(1)} +2 r g_h^{(1)} + r^2 D_B g^B_{(1)} + y_v^{(1)} - 2 r^2 y_h^{(1)} =0\\
    D_A g_h^{(1)} + r^2 \partial_3 g_A^{(1)} + y_A^{(1)} =0
    \end{split}
\end{align}
We expand all the variables into scalar and vector spherical harmonics. 
We have $g_{h/v}^{(1)} = \sum_{l,m} g_{h/v,lm}^{(1)} L_{lm}$, $y_{v/h}^{(1)} = \sum_{l,m} y_{v/h,lm}^{(1)} L_{lm}$ and $g_A^{(1)} = \sum_{I,l,m} g_{I,lm}^{(1)} [L^I_{lm}]_A$.
For simplicity, we will drop the indices $l,m$ in the following equations.
\begin{align}
    \begin{split}
    \label{eq:FirstOrderDEqn}
    4 r^2\partial_3 g_h^{(1)} + 2 r g_h^{(1)} - \sqrt{l(l+1)} r^2 g^{e}_{(1)} + y_v^{(1)} - 2 r^2 y_h^{(1)}=0\\
    \sqrt{l(l+1)} g_h^{(1)}+ r^2 \partial_3 g_e^{(1)} + y_e^{(1)}=0\\
    r^2 \partial_3 g_o^{(1)} + y_o^{(1)} =0
    \end{split}
\end{align}

We start with the odd polarity sector and directly integrate the last equation to obtain the solution for $g_o^{(1)}$:\begin{align}
    g_o^{(1)} = \int \frac{y_o^{(1)}}{r^2}\dd{r} = \int \frac{\sqrt{(l+2)(l-1)}}{r^3} Q^o\dd{r}
\end{align}
where we used the solution for $y_o^{(1)}$:
\begin{align}
    y_o^{(1)}&=\frac{\sqrt{(l+2)(l-1)}}{r} Q^o
\end{align}

For the even polarity we divide the first equation by $r^2$ and take the radial derivative. The radial derivative of $g_e^{(1)}$ is obtained from the second equation. We have the second order differential equation
\begin{align}
    \partial_3 \qty[4 \partial_3 g_h^{(1)} + \frac{1}{r^2}y_v^{(1)} - 2 y_h^{(1)} +\frac{2}{r} g_h^{(1)}] + \frac{\sqrt{l(l+1)} y_e^{(1)}}{r^2} + \frac{l(l+1)}{r^2} g_h^{(1)}=0
\end{align}
Next, we determine the solution of this equation. We first write it in the form 
\begin{align}
    4 \partial_3^2 g_h^{(1)} + \frac{2}{r} \partial_3 g_h^{(1)} + \frac{(l+2)(l-1)}{r^2} g_h^{(1)} = s(r)\,,
\end{align}
where we defined a ``source'' term $s(r)$ as
\begin{align}
    \begin{split}
    s(r) &:= - \partial_3 (r^{-2} y_v^{(1)} - 2 y_h^{(1)}) - \frac{\sqrt{l(l+1)}}{r^2} y_e^{(1)}\\
    &= - \frac{2}{r^2} \partial_r (y_v^{(1)}) + \frac{y_v^{(1)}}{r^3}  + \frac{\sqrt{2(l+2)(l+1)l(l-1)}}{4r^3 \sqrt{r r_s}}(- 2 r (X^e)' + 3 X^e)
    \end{split}
\end{align}

The solution of the linear differential equation consists of a homogeneous solution and a particular solution of the inhomogeneous equation, i.e. the equation with non-vanishing $s(r)$. The homogeneous solution is
\begin{align}
    g_h^{(1)} = C_+ r^{\alpha_+} + C_- r^{\alpha_-}\,,
\end{align}
with the exponents
\begin{equation}
    \alpha_\pm = \frac{1}{4} \pm \frac{i}{4} \sqrt{4l(l+1) - 9}\,.
\end{equation}
The solution of the inhomogeneous problem is given by
\begin{align}
    g_h^{(1)}= r^{\alpha_-} \int \dd{\tilde r} \frac{i \tilde r^{\alpha_+ + 1/2} s(\tilde r)}{2\sqrt{4l(l+1)-9}} - r^{\alpha_+} \int \dd{\tilde r} \frac{i \tilde r^{\alpha_- + 1/2} s(\tilde r)}{2\sqrt{4l(l+1)-9}}
\end{align}
From the asymptotic relation $y_v = O(r^{1/4})$ and $X^e = O(r^{3/4})$, we find that $g_h^{(1)} = O(r^{-3/4})$.

The quantity $g_e^{(1)}$ is determined by equation \eqref{eq:FirstOrderDEqn}.
\begin{align}
    \sqrt{l(l+1)}g_e^{(1)} = 4 \partial_3 g_h^{(1)} + \frac{2}{r} g_h^{(1)} + \frac{\sqrt{2(l+2)(l+1)l(l-1)}}{2 r^2 \sqrt{r r_s}} X^e
\end{align}
It has asymptotic behaviour $g_e^{(1)} = O(r^{-7/4})$.
Then, $g_v^{(1)}$ is given by 
\begin{align}
    \begin{split}
    g_v^{(1)} &= \frac{1}{2 r^2 p_v^{(0)}}\qty(4 r g_h^{(1)} - 2 r^2 \sqrt{l(l+1)} g_e^{(1)} - 2 y_v^{(1)})\\
    &= \frac{- 1}{2 r^2 p_v^{(0)}}\qty( 8 r^2 \partial_3 g_h^{(1)} + \frac{\sqrt{2(l+2)(l+1)l(l-1)}}{\sqrt{r r_s}} X^e + 2 y_v^{(1)})
    \end{split}
\end{align}
with asymptotic behaviour $O(r^{-9/4})$.

\subsection{The second order}
We move on to the second order. 
The spherically symmetric contribution to the stability conditions are
\begin{align}
    \begin{split}
    &8 \pi \partial_3 f_h^{(2)} + 4 \pi f_v^{(2)}(p_v^{(0)} - r^2 p_h^{(0)}) + g_v^{(1)}\cdot (y_v^{(1)} - r^2 y_h^{(1)})\\
    &~~~~~~~~~~~~~~~~~~~~~~~+ 4\pi r^{-2}(p_v^{(2)} - r^2 p_h^{(2)}) - r^{-2}(X_o \cdot Y^o + X_e \cdot Y^e) = 0
    \end{split}\\
    \begin{split}
    &16 \pi r f_h^{(2)} + \frac{2}{\sqrt{2(l+2)(l-1)}}(X^o \cdot g^{(1)}_o + X^e \cdot g^{(1)}_e) - 8\pi r^2 p_v^{(0)} f_v^{(2)} - 2 r^2 g_v^{(1)} \cdot y_v^{(1)}\\
    &~~~~~~~~~~~~~~~~~~~~~~~+ 2 r^{-2}(\frac{p_h^{(0)}}{2}(X^o \cdot X^o + X^e \cdot X^e)  + r^2 (X^o \cdot Y_o + X^e \cdot Y_e) - 4 \pi r^2 p_v^{(2)})=0
    \label{eq:SecondOrderStability2}
    \end{split}
\end{align}

In \cite{I}, we showed that the second order perturbations of the lapse and shift are asymptotically given by
\begin{align}
    f_h^{(2)} \sim \frac{p_v^{(2)}}{2r}, \quad \quad f_v^{(2)} = O(1)
\end{align}
In the following we derive the full expression in the bulk which should match the above asymptotic behaviour.
Equation \eqref{eq:SecondOrderStability2} is solved for $f_v^{(2)}$:
\begin{equation}
\begin{aligned}
    \label{eq:Solutionfv2}
    f_v^{(2)} &= \frac{1}{8 \pi r^2 p_v^{(0)}}\Big[16 \pi r f_h^{(2)} + \frac{2}{\sqrt{2(l+2)(l-1)}}(X^o \cdot g^{(1)}_o + X^e \cdot g^{(1)}_e) - 2 r^2 g_v^{(1)} \cdot y_v^{(1)}\\
    &+ 2 r^{-2}(\frac{p_h^{(0)}}{2}(X^o \cdot X^o + X^e \cdot X^e)  + r^2 (X^o \cdot Y_o + X^e \cdot Y_e) - 4 \pi r^2 p_v^{(2)})\Big]
\end{aligned}
\end{equation}
Then, the first equation gives a differential equation for $f_h^{(2)}$ which is given by
\begin{align}
     &8\pi \partial_3 f_h^{(2)} + \frac{4\pi}{r} f_h^{(2)} + \frac{1}{4 r^2}\Big(\frac{2}{\sqrt{2(l+2)(l-1)}}(X^o \cdot g^{(1)}_o + X^e \cdot g^{(1)}_e) - 2 r^2 g_v^{(1)} \cdot y_v^{(1)}\\
     & + r^{-2}(X^o \cdot X^o + X^e \cdot X^e) p^{(0)}_h - 2 (X^o \cdot Y^o + X^e \cdot Y^e) + 8 \pi p_v^{(2)} - 16 \pi r^2 p_h^{(2)}\Big) + g_v^{(1)}\cdot(y_v^{(1)} - r^2 y_h^{(1)})= 0\nonumber
\end{align}
Next we use the relationship between $p_v^{(2)}$ and $p_h^{(2)}$ to simplify the equation.
We have
\begin{equation}
    p_h^{(2)}= \frac{1}{r} \partial_r p_v^{(2)} - \frac{1}{8 \pi} (Y_o \cdot \partial_r X^o + Y_e \cdot \partial_r X^e)
\end{equation}
Then, the equation becomes
\begin{align}
     &\partial_3 f_h^{(2)} + \frac{1}{2r} f_h^{(2)} - \frac{1}{2r} \partial_r p_v^{(2)} + \frac{1}{4r^2} p_v^{(2)} + \frac{1}{16 \pi} (Y_o \cdot \partial_r X^o + Y_e \cdot \partial_r X^e) + \frac{1}{8 \pi} g_v^{(1)}\cdot(y_v^{(1)} - r^2 y_h^{(1)}) \nonumber \\
     & + \frac{1}{32 \pi r^2}\Big(\frac{2}{\sqrt{2(l+2)(l-1)}}(X^o \cdot g^{(1)}_o + X^e \cdot g^{(1)}_e) - 2 r^2 g_v^{(1)} \cdot y_v^{(1)}+ r^{-2}(X^o \cdot X^o + X^e \cdot X^e) p^{(0)}_h \\
     &~~~~~~~~~~~~~~~~~~~ - 2 (X^o \cdot Y^o + X^e \cdot Y^e) \Big) = 0\nonumber
\end{align}
Let us take the ansatz $f_h^{(2)} = \frac{p_v^{(2)}}{2r} + \tilde f_h^{(2)}$.
Then, we find that $\tilde f_h^{(2)}$ satisfies the equation
\begin{align}
     &\partial_3 \tilde f_h^{(2)} + \frac{1}{2r} \tilde f_h^{(2)} + \frac{1}{16 \pi} (Y_o \cdot \partial_r X^o + Y_e \cdot \partial_r X^e) + \frac{1}{8 \pi} g_v^{(1)}\cdot(y_v^{(1)} - r^2 y_h^{(1)}) \nonumber \\
     & + \frac{1}{32 \pi r^2}\Big(\frac{2}{\sqrt{2(l+2)(l-1)}}(X^o \cdot g^{(1)}_o + X^e \cdot g^{(1)}_e) - 2 r^2 g_v^{(1)} \cdot y_v^{(1)}+ r^{-2}(X^o \cdot X^o + X^e \cdot X^e) p^{(0)}_h \\
     &~~~~~~~~~~~~~~~~~~~ - 2 (X^o \cdot Y^o + X^e \cdot Y^e) \Big) = 0\nonumber
\end{align}

The new terms give the integral
\begin{equation}
    \begin{aligned}
        \tilde f_h^{(2)} &= \sqrt{\frac{r_s}{r}} C - \frac{\sqrt{r r_s}}{32 \pi r} \int \frac{1}{r \sqrt{r r_s}}\Big(\frac{2}{\sqrt{2(l+2)(l-1)}}(X^o \cdot g^{(1)}_o + X^e \cdot g^{(1)}_e) + 2 r^2 g_v^{(1)} \cdot (y_v^{(1)} - 2 r^2 y_h^{(1)})\\
        &~~~~~~~~~+ r^{-2}(X^o \cdot X^o + X^e \cdot X^e) p^{(0)}_h - 2 (X^o \cdot Y^o + X^e \cdot Y^e)  +  2r^2 (Y_o \cdot \partial_r X^o + Y_e \cdot \partial_r X^e) \Big) \dd r
    \end{aligned}
\end{equation}
where $C$ is an integration constant that can be reabsorbed into the zeroth order solution $f_h^{(0)}$.
Inserting the asymptotic solution for the momenta $y$ and the test functions $g$, we find that the integral behaves as $O(r^{-1/2})$. 
Therefore, at infinity, the correction $\tilde f_h^{(2)}$ vanishes up to a constant that can be reabsorbed into the zeroth order solution.
Therefore, our result is consistent with the asymptotic behaviour derived in \cite{I}.

Finally, we have the solution for the perturbed lapse to second order:
\begin{equation}
\begin{aligned}
    f_v^{(2)} &= \frac{1}{8 \pi r^2 p_v^{(0)}}\Big[16 \pi r \tilde f_h^{(2)} + \frac{2}{\sqrt{2(l+2)(l-1)}}(X^o \cdot g^{(1)}_o + X^e \cdot g^{(1)}_e) - 2 r^2 g_v^{(1)} \cdot y_v^{(1)}\\
    &+ 2 r^{-2}(\frac{p_h^{(0)}}{2}(X^o \cdot X^o + X^e \cdot X^e)  + r^2 (X^o \cdot Y_o + X^e \cdot Y_e))\Big]
\end{aligned}
\end{equation}
Inserting the leading order asymptotic behaviour, we find 
\begin{equation}
    f_v^{(2)} \sim \frac{2}{r p_v^{(0)}} \tilde f_h^{(2)} \sim \frac{p_v^{(2)}}{r^2 p_v^{(0)}} 
\end{equation}
The asymptotic behaviour of $p_v^{(2)}$ is $O(1)$ and thus, $f_v^{(2)}$ can be neglected compared to $f_v^{(0)} = r^{-2}$.
Hence, our result is consistent with the asymptotic behaviour derived earlier.

\section{Penrose diagrams in deformed Schwarzschild spacetimes}
\label{sec:Penrose}

In the previous section, we used the gauge fixing conditions to derive differential equations fixing the lapse function $S_\bot$ and shift vector $S^i$.
Within our perturbative analysis, we explicitly computed these objects up to second order in perturbation theory.
Thus, we have access to the full four-dimensional metric in terms of the physical degrees of freedom $(X,Y)$ and $r_s=2M$:
\begin{equation}
    \dd s^2 = - (S_\bot^2 - (S^3)^2) \dd t^2 + 2 S^3 \dd t \dd r + \dd r^2 + (r^2 \Omega_{AB} + X_{AB}) (S^A \dd t + \dd x^A)(S^B \dd t + \dd x^B)
    \label{eq:Metric4D}
\end{equation}
where $S_\bot$, $S^3$ and $S^A$ are perturbative expressions in terms of $(X,Y)$.

The metric in \eqref{eq:Metric4D} can be analysed using standard techniques of classical and quantum general relativity.
Classically, one can study Penrose diagrams of the perturbed metric and look for modifications of the causal structure due to the perturbations.
In section \ref{sec:ApparentHorizon}, we saw that the location of the apparent horizon is shifted for non-trivial perturbations.
This could influence the horizons in the corresponding Penrose diagram. 
In the quantum theory the four dimensional metric expressed in terms of $X$ and $Y$ variables can be viewed as an operator acting on the Hilbert space.
Given some states in the Hilbert space, expectation values for the metric can be computed and the resulting ``effective'' metric might lead to quantum corrections
of the Penrose diagram.

In the following, we aim to derive Penrose diagrams for metrics of the above form.
We begin by recalling the construction of Penrose diagrams for spherically symmetric
metrics such as the metric confined to zeroth order in perturbation theory. Thus
we suppress the angular dimension, ignore the angular components of the metric $\dd x^A = 0$ 
which do not contribute to radial null geodesics and focus only on the $(t,r)$ part.
Furthermore, we simplify to only spherically symmetric functions $S_\bot$ and $S^3$. 
Then, the metric reads
\begin{equation}
    \dd s^2 = - S_\bot^2 \dd t^2 + (S^3 \dd t + \dd r)^2
\end{equation}
In the previous section we computed to zeroth order $S_\bot = 1$ and $S^3 = \sqrt{\frac{r_s}{r}}$.
The out-/ingoing radial null geodesics $r_\pm$ satisfy the differential equation
\begin{equation}
    \dot r_\pm = \dv{r_\pm}{t} = - S^3 \pm \abs{S_\bot}
    \label{eq:NullGeodesicSym}
\end{equation}
We introduce two functions $U,V:(t,r) \to \mathbb{R}$ such that they are constant along the solutions, i.e.
$U(t,r_+(t))$ and $V(t,r_-(t))$ are constant.
In other words, $U$ satisfies the differential equation
\begin{equation}
    \dv{U}{t} = \pdv{U}{t} + \pdv{U}{r} \dot r_+ = 0
    \label{eq:pdeU}
\end{equation}
and similarly for $V$. By construction we have
\begin{equation}
    \dd U = \pdv{U}{t} \dd t + \pdv{U}{r} \dd r = - \pdv{U}{r}\dot r_+ \dd t + \pdv{U}{r} \dd r
\end{equation}
Then, we compute
\begin{align}
    \begin{aligned}
        \dd U \dd V &= \pdv{U}{r} \pdv{V}{r}\qty(\dot r_+ \dot r_- \dd t^2- (\dot r_+ + \dot r_-) \dd t \dd r + \dd r^2)\\
        &= \pdv{U}{r} \pdv{V}{r}\qty(- \qty(S_\bot^2 - (S^3)^2) \dd t^2 + 2 S^3 \dd t \dd r + \dd r^2)\\
        &= \pdv{U}{r} \pdv{V}{r} \dd s^2
    \end{aligned}
\end{align}
Therefore, provided the prefactor is non-zero, we can write the original metric as
\begin{equation}
    \dd s^2 = \qty(\pdv{U}{r} \pdv{V}{r})^{-1} \dd U \dd V
\end{equation}
Notice that $U,V$ will be good coordinates locally provided that the Jacobian relating them to the original coordinates $(t,r)$ is non-singular.
In particular, we need to have
\begin{align}
    \begin{split}
    \dd U \wedge \dd V &= (\partial_t U \partial_r V - \partial_r V \partial_t V )\dd t \wedge \dd r\\
    &=\partial_r U \partial_r V (\dot r_+ - \dot r_-) \dd t \wedge \dd r \neq 0
    \end{split}
\end{align}
Thus, we require $\dot r_+ - \dot r_- = 2 S_\bot \neq 0$ and $\partial_r U \partial_r V \neq 0$.

The functions $U$ and $V$ are only defined up to a reparametrisation $U \to F(U)$. 
It is easy to see, that if $U$ satisfies equation \eqref{eq:pdeU} then also $F(U)$ does.
In order to find explicit expressions for $U$, we write the functions $U$ and $V$ as
\begin{equation}
    U = - e^{-(t + f_U(r))/(2 r_s)}, \qquad V = e^{(t + f_V(r))/(2 r_s)}
\end{equation}
This particular parametrization is motivated by the Kruskal coordinate system for the Schwarzschild black hole.
For this choice, $f_U$ and $f_V$ need to satisfy the equation
\begin{equation}
    \partial_r f_U = - \frac{1}{\dot r_+}, \qquad \partial_r f_V = - \frac{1}{\dot r_-}
\end{equation}
In the case of the Schwarzschild black hole in Gullstrand-Painlevé coordinates we have $S_\bot = 1$ and $S^3 = \sqrt{\frac{r_s}{r}}$.
Then, the equations can be integrated explicitly:
\begin{align}
    \begin{aligned}
        f_U &= - r - 2 \sqrt{r r_s} - 2 r_s \log(\sqrt{\frac{r}{r_s}} - 1)\\
        f_V &= + r - 2 \sqrt{r r_s} + 2 r_s \log(\sqrt{\frac{r}{r_s}}+1)
    \end{aligned}
\end{align}
We find a relation between $r$ and the product of $U$ and $V$:
\begin{equation}
    U V = - e^{(f_V - f_U)/(2 r_s)} = e^{r/r_s}\qty(1 - \frac{r}{r_s})
    \label{eq:UVrRelation}
\end{equation}
For the computation of the metric, we compute
\begin{align}
    \begin{aligned}
    \partial_r U \partial_r V &= \frac{1}{4 r_s^2} e^{(f_V - f_U)/(2 r_s)} \frac{1}{\dot r_+ \dot r_-} =- \frac{e^{r/r_s}}{4 r_s^2}\qty(1 - \frac{r}{r_s}) \frac{1}{\frac{r_s}{r} - 1} = - \frac{e^{r/r_s}r}{4 r_s^3}
    \end{aligned}
\end{align}
and find the metric in $U$ and $V$ coordinates:
\begin{equation}
    \dd s^2 = - \frac{4 r_s^3}{r} e^{- r/r_s} \dd U \dd V
\end{equation}
This is the Schwarzschild metric in Kruskal coordinates where $r$ is expressed in terms of $U,V$ through equation \eqref{eq:UVrRelation}.
Compactifying the coordinates $U,V$ using the inverse tangent, we obtain the Penrose diagram for the Schwarzschild black hole.\\
\\
The Penrose diagram corresponding to our spherically symmetric spacetime at zeroth order in 
the perturbations is now standard. To incorporate the perturbations up to second order we
can now use the formulae of the previous sections in order to express the full spacetime metric 
up to second order in the fields $X,Y$. In order to construct a Penrose diagram for the 
perturbed spacetime which is no longer spherically symmetric we face the problem that the 
double null coordinates that one uses in the construction of Penrose diagrams, which 
suppress the angular dimensions, become 
now direction dependent corresponding to the initial tangent of radial geodesics at large radius.
Thus, to obtain a qualitative picture of the spacetime which is also no longer stationary, 
one would need to construct several, Penrose diagrams corresponding to an equidistant 
distribution (with respect to the round sphere metric) of directions on the sphere. To avoid 
that complication, as a crude approximation we content ourselves by generating a spherically
symmetric metric using angular averaging over the sphere. We can then use the above construction
for spherically symmetric metrics. 
This captures at least some aspects of of the perturbed Penrose diagram. We will not do so in all detail but confine 
ourselves to showing qualitatively that indeed the effective apparent horizon radius is shifted.    
Note that all computations are entirely classical. However, to capture quantum effects one would simply 
redo the computations performed below for the corresponding expectation values rather than the 
classical objects.\\
\\
In section \ref{sec:LapseShift} we derived expressions for the perturbed lapse and shift up to second order in the perturbations. 
Averaging the results over the sphere, we observe that the second order corrections contribute to the spherically symmetric part of lapse and shift. 
Thus, the perturbations generate symmetric corrections to the spherically symmetric metric. 
Explicitly, we find 
\begin{align}
\begin{aligned}
    S^\mathrm{sym}_\bot &= \frac{1}{4 \pi} \int \dd \Omega S_\bot = \qty(r^2  - \frac{1}{16 \pi r^2} \sum_I X^I \cdot X_I) f_v^{(0)} + r^2 f_v^{(2)}\\
    S^3_\mathrm{sym} &= \frac{1}{4\pi} \int \dd \Omega S^3 =f_h^{(0)} + f_h^{(2)}
\end{aligned}
\end{align}

As before, null geodesics satisfy equation \eqref{eq:NullGeodesicSym}. 
Expanding lapse and shift into the corrected spherically symmetric contributions, we find the equation
\begin{equation}
    \begin{aligned}
    \dot r_\pm &= - S^3_\mathrm{sym} \pm S^\mathrm{sym}_\bot = - \sqrt{\frac{r_s}{r}} \pm 1 - f_h^{(2)} \pm \qty(r^2 f_v^{(2)} - \frac{1}{16 \pi r^4} \sum_I X^I \cdot X_I )\\
    &= - \sqrt{\frac{r_s}{r}} \pm 1 - \frac{p_v^{(2)}}{2r} + \qty(-1 \pm \sqrt{\frac{r}{r_s}})\tilde f_h^{(2)} \pm \frac{1}{8 \pi p_v^{(0)}}\Big[\frac{2}{\sqrt{2(l+2)(l-1)}}(X^o \cdot g^{(1)}_o + X^e \cdot g^{(1)}_e)\\
    &- 2 r^2 g_v^{(1)} \cdot y_v^{(1)} + 2 (X^o \cdot Y_o + X^e \cdot Y_e)) \Big]
    \end{aligned}
\end{equation}
with
\begin{equation}
    \begin{aligned}
        \tilde f_h^{(2)} &= - \frac{\sqrt{r r_s}}{32 \pi r} \int \frac{1}{r \sqrt{r r_s}}\Big(\frac{2}{\sqrt{2(l+2)(l-1)}}(X^o \cdot g^{(1)}_o + X^e \cdot g^{(1)}_e) + 2 r^2 g_v^{(1)} \cdot (y_v^{(1)} - 2 r^2 y_h^{(1)})\\
        &~~~~~~~~~+ r^{-2}(X^o \cdot X^o + X^e \cdot X^e) p^{(0)}_h - 2 (X^o \cdot Y^o + X^e \cdot Y^e)  +  2r^2 (Y_o \cdot \partial_r X^o + Y_e \cdot \partial_r X^e) \Big) \dd r
    \end{aligned}
\end{equation}

Then, the zero of $\dot r_+$ is shifted indicating a change in the position of the apparent horizon. To see this,
assuming the ansatz $r = r_s + \delta r$, where $\delta r$ is second order in the perturbations, we find the condition
\begin{equation}
    0 = \dot r_+ \sim \frac{1}{2 r_s} \delta r - \frac{p_v^{(2)}(r_s)}{2 r_s} + \frac{1}{16 \pi r_s}\qty[\sum_I 2 X^I \cdot \qty(Y_I + \frac{1}{\sqrt{2(l+2)(l-1)}} g_I^{(1)}) - 2 r_s^2 g_v^{(1)} \cdot y_v^{(1)}]
\end{equation}
which can be solved for the correction $\delta r$:
\begin{equation}
    \delta r = p_v^{(2)}(r_s) - \frac{1}{8 \pi}\qty[\sum_I 2 X^I \cdot \qty(Y_I + \frac{1}{\sqrt{2(l+2)(l-1)}} g_I^{(1)}) - 2 r_s^2 g_v^{(1)} \cdot y_v^{(1)}]_{r = r_s}
\end{equation}
Thus, the position of the zero of $\dot r_+$ is shifted in this spherically symmetric averaged spacetime. 
The relation to the apparent horizon computed in section \ref{sec:ApparentHorizon} can be worked out using 
the formulae provided in the previous sections, the exact formula will be provided in future contributions 
to this series.

\section{Conclusion}
\label{sec:Conclusion}

The physics of evaporating black holes, especially the final phase, where black holes have small mass, is poorly understood.
In this regime, the semi-classical approximation is expected to break down. 
According to the semi-classical approximation the emitted power increases and eventually the change of the black hole spacetime strongly contradicts the adiabatic assumption of a static background geometry.
For massive and super-massive black holes, the observable signal from Hawking radiation is very weak.
Thus, the final phase is particularly important for observations of Hawking radiation.
In our first principle approach we intend to shed new light on this critical phase of Hawking evaporation. 
Using a gauge-invariant formulation including back reaction valid to any order we obtained the dynamics of the reduced phase space explicitly to second 
and third order.
In this manuscript we study interesting observables on the reduced phase space, the area of the apparent horizon and the quasi-local mass. 

We computed the shape of the apparent horizon to second order in the perturbations and obtained an explicit expression for the area of the apparent horizon.
The square root of the area defines a dynamical notion of mass for the black hole. 
We showed that the change of this quasi-local mass under time evolution 
with respect to Painleve-Gullstrand foliations is positive in the classical theory and hence black holes will not evaporate 
classically. 
Due to mechanisms familiar from the violation of energy inequalities in quantum field theories, 
a modification of the classical area inequalities in the quantum theory and a decrease of the quasi-local mass is conceivable.

We saw that the zeroth order contribution to the area of the apparent horizon without matter 
is just the area of a sphere with the Schwarzschild (vacuum) or Reissner-Nordst\o{}m radius
as coordinate radius. To second order, the apparent horizon 
receives non spherically symmetric corrections from the perturbation fields and likewise its 
area depends not only on the mass but also on second order contributions from the 
gravitational perturbations. In case of gravity and electromagnetic matter the only 
symmetric degrees of freedom are mass and charge.
Adding additional matter (e.g. scalar fields, neutrinos, \dots), we will obtain in addition 
spherically symmetric fields with corresponding non-trivial background dynamics 
and see even more interesting corrections to the shape and area of the apparent horizon.

In the future, we plan to study the quantisation of the reduced phase space and the physical Hamiltonian.
An effective description will be available by considering expectation values for both the area of the apparent horizon and the quasi-local mass in 
physically interesting states.
For the perturbations the structure of the physical Hamiltonian suggests a Fock representation.
The mode functions are in principle explicitly known in terms of Heun functions \cite{Perlick1,Perlick2}.
However, interesting new mathematical challenges arise when gluing past ingoing and
future outgoing Gullstrand-Painleve spacetimes to obtain a globally hyperbolic spacetime \cite{1} 
which will be subject of other contributions to this series.

\appendix

\section{Boundary Term Even Polarity Perturbations Coupled to Electromagnetic Matter}
\label{sec:EvenParityBdryEM}

The boundary term $B_e$ is given by
\small
\begin{align}
    \label{eq:BeEM}
    B_e &= - \frac{1}{r^2}X^e \cdot \partial_r X^e + \frac{r \Delta}{l(l+1) \Lambda}  \partial_r y_v^{(1)} \cdot \partial_r y_v^{(1)} + \frac{2 g^2 \xi \Delta}{\Lambda p_v^{(0)} \sqrt{l(l+1)}} \partial_r y_v^{(1)} \cdot A + \sqrt{\frac{(l+2)(l-1)}{2 l(l+1)}} \frac{2 \Delta}{\Lambda p_v^{(0)}} \partial_r y_v^{(1)} \cdot X^e\\
    & + \frac{g^4 \xi ^4 - 2 (l+2)(l-1) g^2 \xi ^2 r^2+2 r r_s \left(-3 g^2 \xi ^2+4 \left(l^2+l-1\right) r^2+4 r r_s\right)}{2 r^2 \Lambda l(l+1) (p_v^{(0)})^2} y_v^{(1)} \cdot \partial_r y_v^{(1)}\nonumber\\
    &+ \frac{\sqrt{2(l+2)(l-1)}g^2 \xi \Delta(2-\Delta)}{\Lambda r (p_v^{(0)})^2} X^e \cdot A + \frac{\Delta (2 - \Delta) g^4 \xi^2}{r \Lambda p_v^{(0)}} A \cdot A\nonumber\\
    &+\frac{1}{32 r^7 \Lambda (p_v^{(0)})^2}\Big[-g^6 \xi ^6+20 g^4 \xi ^4 r^2-20 g^2 (l+2)(l-1) \xi ^2 r^4 + 16 (l+2)(l-1) r^6\nonumber\\
    &~~~~~~~~~+r r_s \left(7 g^4 \xi ^4-4 r_s \left(3 g^2 \xi ^2 r+4 \left(l^2+l-17\right)
    r^3\right)+4 g^2 \left(l^2+l-37\right) \xi ^2 r^2+80 \left(l^2+l-2\right) r^4\right)\Big] X^e \cdot X^e\nonumber\\
    &+ \frac{g^2 \xi}{16 \sqrt{l(l+1)} r^7 \Lambda^2 (p_v^{(0)})^2} \Big[g^8 \xi ^8-g^6 \left(l^2+l+2\right) \xi ^6 r^2+2 g^4 (l-1) (l+2) \left(l^2+l+8\right) \xi ^4 r^4\nonumber\\
    &~~~~~~~~~-8 g^2 \left(l^2+l-2\right)^2 \xi ^2
    r^6+2 r r_s \left(-6 g^6 \xi ^6+g^4 (5 l (l+1)+14) \xi ^4 r^2-8 g^2 \left(l (l+1) \left(l^2+l+5\right)-12\right) \xi ^2 r^4\right.\nonumber\\
    &~~~~~~~~~+4 r r_s \left(6 g^4 \xi^4+4 r_s \left(\left(l^2+l+7\right) r^3-2 g^2 \xi ^2 r\right)-g^2 (4 l (l+1)+17) \xi ^2 r^2+4 \left(l (l+1) \left(l^2+l+4\right)-9\right)
    r^4\right)\nonumber\\
    &~~~~~~~~~\left.+16 (l-1) (l+2) \left(l^2+l-1\right) r^6\right)] y_v^{(1)} \cdot A\nonumber\\
    &+ \sqrt{\frac{(l+2)(l-1)}{2l(l+1)}}\frac{1}{16 r^7 \Lambda^2 (p_v^{(0)})^3}\Big[2 g^8 \xi ^8-2 g^6 \left(l^2+l-2\right) \xi ^6 r^2+2 g^4 (l-1) (l+2) \left(l^2+l+6\right) \xi ^4 r^4\nonumber\\
    &~~~~~~~~~-8 g^2 \left(l^2+l-2\right)^2 \xi ^2 r^6+r r_s
    \left(-27 g^6 \xi ^6+2 g^4 (11 l (l+1)-20) \xi ^4 r^2-16 g^2 \left(l (l+1) \left(l^2+l+3\right)-8\right) \xi ^2 r^4\right.\nonumber\\
    &~~~~~~~~~+4 r r_s \left(33 g^4 \xi ^4+4 r
    r_s \left(-17 g^2 \xi ^2+6 \left(l^2+l-1\right) r^2+12 r r_s\right)-10 g^2 (2 l (l+1)-3) \xi ^2 r^2\right.\nonumber\\
    &~~~~~~~~~\left.\left.+8 \left(l (l+1) \left(l^2+l+2\right)-5\right)
    r^4\right)+32 (l-1) (l+2) \left(l^2+l-1\right) r^6\right)] y_v^{(1)} \cdot X^e\nonumber\\
    &+\frac{1}{64 l(l+1) r^9 \Lambda^3 (p_v^{(0)})^4}\Big[g^4 \xi ^4 \left(-g^8 \xi ^8+g^6 \left(l^2+l+6\right) \xi ^6 r^2-4 g^4 (l-1) (l+2) \left(l^2+l+7\right) \xi ^4 r^4\right.\nonumber\\
    &~~~~~~~~~\left.+20 g^2 \left(l^2+l-2\right)^2 \xi^2 r^6-8 \left(l^2+l-2\right)^3 r^8\right)\nonumber\\
    &~~~~~~~~~+r r_s \left(g^2 \xi ^2 \left(16 g^8 \xi ^8-11 g^6 \left(l^2+l+10\right) \xi ^6 r^2+2 g^4 (l-1) (l+2) (23 l (l+1)+162) \xi ^4 r^4\right.\right.\nonumber\\
    &~~~~~~~~~\left.+4 g^2 \left(l^2+l-36\right) \left(l^2+l-2\right)^2 \xi ^2 r^6+32 \left(l^2+l-2\right)^3 r^8\right)\nonumber\\
    &~~~~~~~~~+2 r r_s \left(-48 g^8 \xi^8+g^6 (17 l (l+1)+394) \xi ^6 r^2-4 g^4 (l (l+1) (22 l (l+1)+123)-342) \xi ^4 r^4\right.\nonumber\\
    &~~~~~~~~~-8 g^2 (l-1) (l+2) (l (l+1) (2 l (l+1)-21)+42) \xi ^2 r^6+2 r
    r_s \left(64 g^6 \xi ^6+3 g^4 \left(l^2+l-224\right) \xi ^4 r^2\right.\nonumber\\
    &~~~~~~~~~+8 g^2 \left(7 l (l+1) \left(l^2+l+8\right)-152\right) \xi ^2 r^4-4 r_s \left(16 g^4 \xi ^4 r+3 g^2 (4 l (l+1)-87) \xi ^2 r^3+36 (2 l (l+1)-5) r^5\right)\nonumber\\
    &~~~~~~~~~\left.\left.\left.+16 (l-1) (l+2) \left(l (l+1) \left(l^2+l-3\right)+8\right) r^6+48 \left(l^2+l-11\right) r^4 r_s^2\right)+182 \left(l^2+l-2\right)^2 r^8\right)\right)\Big]y_v^{(1)} \cdot y_v^{(1)}\nonumber
\end{align}
\normalsize
In this equations we defined $A = - \int Y^e_{lm} \dd r$.
In addition, we used the quantity $\Delta$:
\begin{equation}
    \Delta = 1 - \frac{r_s}{r} - \frac{g^2 \xi^2}{r^2}
\end{equation}

\section{Third Order Apparent Horizon}
\label{sec:ThirdOrder}

In this appendix we extend the computations in section \ref{sec:ApparentHorizon} to third order corrections.
In a recent publication \cite{V}, we continued the perturbative expansion of the physical Hamiltonian to third order.
In the future these results together with the results in this appendix will give us access to higher order effects.

We start by computing the expansion of the outward null expansion in equation \eqref{eq:ApparentHorizon} to third order.
The momentum terms to third order are given by
\begin{align}
    - s_i s_j W^{ij} &= \sqrt{\Omega}\Big(- (1 - m^{AB} D_A \rho D_B \rho) (p_v + y_v) + D_A \rho y^A - \frac{1}{2} D_A \rho D_B \rho \Omega^{AB} (p_h + y_h) - D_A \rho D_B \rho Y^{AB}\Big)\nonumber
\end{align}
Next, we find
\begin{align}
    \sqrt{m}m^{33} s_3 &\sim r^2 \sqrt{\Omega} \qty(1 - \frac{1}{4 r^4} X^{AB} X_{AB} - \frac{1}{2} m^{AB} D_A \rho D_B \rho)\\
    \sqrt{m}m^{AB} s_B &\sim \sqrt{\Omega} \qty(- D^A \rho  + \frac{1}{r^2}X^{AB} D_B \rho + \frac{1}{4 r^4}X^{BC} X_{BC} D^A \rho - \frac{1}{r^4} X^{AC} X^{B}{}_C D_B \rho + \frac{1}{2 r^2} \Omega^{BC} D_B \rho D_C \rho D^A\rho)\nonumber\\
    &= \sqrt{\Omega} \qty(- D^A \rho  + \frac{1}{r^2}X^{AB} D_B \rho - \frac{1}{4 r^4}X^{BC} X_{BC} D^A \rho + \frac{1}{2 r^2} \Omega^{BC} D_B \rho D_C \rho D^A\rho)
\end{align}
In the last step we used the identity $X^{AC} X^B{}_C = \frac{1}{2} \Omega^{AB} X^{CD} X_{CD}$ (see \cite{V}).
Taking the corresponding derivatives, we obtain
\begin{align}
    \partial_3 (\sqrt{m} m^{33} s_3) &\sim \sqrt{\Omega} \qty(2r - \frac{1}{2 r} X^{AB} \partial_r (r^{-1} X_{AB}) + \frac{1}{2} \partial_r \qty(r^{-2} X^{AB}) D_A \rho D_B \rho)\\
    \partial_A (\sqrt{m} m^{AB} s_B) &\sim \sqrt{\Omega} D_A\qty(- D^A \rho  + \frac{1}{r^2}X^{AB} D_B \rho - \frac{1}{4 r^4}X^{BC} X_{BC} D^A \rho + \frac{1}{2 r^2} \Omega^{BC} D_B \rho D_C \rho D^A\rho)
\end{align}
Hence, the outward expansion $\theta_+$ equals
\begin{align}
    \begin{aligned}
    \frac{\sqrt{m}}{\sqrt{\Omega}} \theta_+ &= - (1 - \frac{1}{2} m^{AB} D_A \rho D_B \rho) (p_v + y_v) + D_A \rho y^A - \frac{1}{2} D_A \rho D_B \rho \Omega^{AB} (p_h + y_h) - D_A \rho D_B \rho Y^{AB}\\
    &+ 2r - \frac{1}{2 r} X^{AB} \partial_r (r^{-1} X_{AB}) + \frac{1}{2} \partial_r \qty(r^{-2}X^{AB}) D_A \rho D_B \rho\\
    &+ D_A\qty(- D^A \rho  + \frac{1}{r^2}X^{AB} D_B \rho - \frac{1}{4 r^4}X^{BC} X_{BC} D^A \rho - \frac{1}{2 r^2} \Omega^{BC} D_B \rho D_C \rho D^A\rho)
    \end{aligned}
\end{align}
In this expression, we insert a perturbative expansion of the function $\rho$ as $\rho = \rho^{(0)} + \dots + \rho^{(3)}$, where $\rho^{(i)}$ is of order $i$ in the perturbations.
As in section \ref{sec:ApparentHorizon}, the indirect dependence of $\rho$ is treated using a Taylor series expansion around the spherically symmetric solution $\rho^{(0)}$.
Then, the symmetric component $\rho^{(3)}_{00}$ is the solution of
\begin{align}
    \begin{aligned}
    &\rho^{(3)}_{00} - \sqrt{4 \pi} p_v^{(3)} + \frac{1}{2r_s} \rho^{(1)} \cdot \rho^{(2)} - \frac{1}{8 r_s^2} \ev{(\rho^{(1)})^3} - \partial_r y_v^{(2)} \cdot \rho^{(1)} - \partial_r y_v^{(1)} \cdot \rho^{(2)} - \frac{1}{2} \ev{\partial_r^2 y_v^{(1)} (\rho^{(1)})^2}\\
    &- \frac{3}{4r_s^2} \ev{D^A \rho^{(1)} D_A \rho^{(1)} \rho^{(1)}} + \frac{1}{r_s} \ev{D^A \rho^{(2)} D_A \rho^{(1)}} + \ev{D_A \rho^{(1)} y^A_{(2)}} + \ev{D_A \rho^{(2)}y^A_{(1)}}\\
    &+ \ev{D_A \rho^{(1)}\partial_r y^A_{(1)} \rho^{(1)}} - \frac{1}{2 r_s^2} \ev{\partial_r X^{AB} \partial_r X_{AB} \rho^{(1)}} - \frac{1}{2 r_s^2} \ev{X^{AB} \partial_r^2 X_{AB} \rho^{(1)}} + \frac{3}{2r_s^3} \ev{X^{AB} \partial_r X_{AB} \rho^{(1)}}\\
    & - \frac{3}{2 r_s^4}\ev{X^{AB} X_{AB} \rho^{(1)}} + \frac{1}{2 r_s^2} \ev{D^A \rho^{(1)}D_A \rho^{(1)} \qty(y_v^{(1)} - r_s^2 y_h^{(1)})}\\
    &+ \frac{1}{2} \ev{\qty(r_s^{-2}\partial_r X^{AB} - 4 r_s^{-3} X^{AB} - 2 Y^{AB}) D_A \rho^{(1)} D_B \rho^{(1)}} = 0
    \end{aligned}
\end{align}
For any function of the angles $f$, we defined its average over the sphere as
\begin{equation}
    \langle f \rangle = \int_{S^2} f \dd \Omega
\end{equation}

Next, we compute the area of the apparent horizon to third order. 
We start by computing the induced metric on the horizon and then integrate the associated volume element over the sphere.
The induced metric is 
\begin{align}
    \begin{aligned}
    h_{AB} &= Y^i_{(A)} Y^j_{(B)} m_{ij} = \rho^2 \Omega_{AB} + X_{AB} + D_A \rho D_B \rho\\
    &\sim \qty(r_s^2 + 2 r_s(\rho^{(1)} + \rho^{(2)} + \rho^{(3)}) + (\rho^{(1)})^2 + 2 \rho^{(1)} \rho^{(2)})\Omega_{AB} + D_A \rho^{(1)} D_B \rho^{(1)} + 2 D_{(A}\rho^{(2)} D_{B)}\rho^{(1)}\\
    &+ X_{AB} + \partial_r X_{AB} (\rho^{(1)} + \rho^{(2)}) + \frac{1}{2} \partial_r^2 X_{AB} (\rho^{(1)})^2 
    \end{aligned}
\end{align}
Consider a general tensor $T_{AB}$. In two dimensions, we have the identity
\begin{equation}
    \det(r_s^2 \Omega + T) = r_s^4 \det \Omega + r_s^2 \det \Omega \Omega^{AB} T_{AB} + \det T
\end{equation}
with
\begin{equation}
    \det T = \frac{1}{2}\det \Omega\qty((\Omega^{AB} T_{AB})^2 - \Omega^{AB} \Omega^{CD} T_{AC} T_{BD}) 
\end{equation}
Then, we find for the square root of the determinant to third order:
\begin{align}
    &\sqrt{\det(r_s^2 \Omega + T)} = r_s^2 \sqrt{\det \Omega}\sqrt{1 + \frac{1}{r_s^2} \Omega^{AB} T_{AB} + \frac{1}{r_s^4}\frac{\det T}{\det \Omega}}\\
    &\sim r_s^2 \sqrt{\det \Omega}\qty(1 + \frac{1}{2 r_s^2} \Omega^{AB} T_{AB} + \frac{1}{2r_s^4}\frac{\det T}{\det \Omega} - \frac{1}{8 r_s^4} (\Omega^{AB}T_{AB})^2 - \frac{1}{4 r_s^6}\Omega^{AB}T_{AB} \frac{\det T}{\det \Omega} + \frac{1}{16 r_s^6}  (\Omega^{AB}T_{AB})^3)\nonumber
\end{align}
where we used that $\det T$ is of order two if $T$ is of order $1$ in the perturbations.
Hence, we find
\begin{align}
    &\sqrt{\det h} = r_s^2 \sqrt{\det \Omega}\Big(1 + \frac{2}{r_s} \rho^{(1)} + \frac{2}{r_s} \rho^{(2)} + \frac{1}{r_s^2} (\rho^{(1)})^2 +\frac{1}{2 r_s^2}D^A \rho^{(1)}D_A \rho^{(1)} - \frac{1}{4 r_s^4} X^{AB} X_{AB}\\
    &+ \frac{2}{r_s} \rho^{(3)} + \frac{2}{r_s^2}\rho^{(1)} \rho^{(2)} + \frac{1}{r_s^2}D^A \rho^{(1)} D_A \rho^{(2)} - \frac{1}{2 r_s^4} X^{AB}  (D_A \rho^{(1)} D_B \rho^{(1)} + \partial_r X_{AB} \rho^{(1)}) + \frac{1}{2 r_s^5} \rho^{(1)} X^{AB} X_{AB}\Big)\nonumber
\end{align}
Integrating this over the sphere gives the area:
\begin{align}
    \begin{aligned}
        A &= 4 \pi r_s^2 + 2 r_s (\rho^{(2)}_{00} + \rho^{(3)}_{00}) + \frac{l(l+1) + 2}{2}\rho^{(1)} \cdot \rho^{(1)} - \frac{1}{4 r_s^2} \sum_I X^I \cdot X^I + (l(l+1) + 2) \rho^{(1)} \cdot \rho^{(2)}\\
        &- \frac{1}{2 r_s^2} \langle X^{AB} D_A \rho^{(1)} D_B \rho^{(1)}\rangle - \frac{1}{2 r_s^2} \langle X^{AB} \partial_r X_{AB} \rho^{(1)}\rangle + \frac{1}{2 r_s^3} \langle \rho^{(1)} X^{AB} X_{AB} \rangle
    \end{aligned}
\end{align}
The second order was already computed in the main text. Let us focus on the third order contributions:
\begin{align}
    \begin{split}
    A^{(3)} &= 2 r_s \rho^{(3)}_{00} + (l(l+1)+2) \rho^{(1)} \cdot \rho^{(2)} + \frac{2}{r_s} \langle (\rho^{(1)})^3 \rangle + \frac{1}{2 r_s^3} \langle \rho^{(1)} X^{AB} X_{AB} \rangle\\
    & - \frac{1}{2 r_s^2}\qty( \langle X^{AB} D_A \rho^{(1)} D_B \rho^{(1)}\rangle + \langle X^{AB} \partial_r X_{AB} \rho^{(1)}\rangle)
    \end{split}
\end{align}
In this expression we need to insert the solution for $\rho_{00}^{(3)}$ and use the solutions for the constraints up to third order in \cite{V}.

\providecommand{\href}[2]{#2}\begingroup\raggedright\endgroup

\end{document}